\documentclass{aa}
\pdfoutput=1
\usepackage[varg]{txfonts}
\usepackage[caption = false]{subfig}
\usepackage{graphicx}
\usepackage{amsmath}
\usepackage{amssymb}
\usepackage{color}
\usepackage[colorinlistoftodos]{todonotes}
\usepackage{bm}

\usepackage{natbib,twoopt}
\usepackage[breaklinks=true]{hyperref} 
\bibpunct{(}{)}{;}{a}{}{,}             
\makeatletter
  \newcommandtwoopt{\citeads}[3][][]{\href{http://adsabs.harvard.edu/abs/#3}%
    {\def\hyper@linkstart##1##2{}%
     \let\hyper@linkend\@empty\citealp[#1][#2]{#3}}}
  \newcommandtwoopt{\citepads}[3][][]{\href{http://adsabs.harvard.edu/abs/#3}%
    {\def\hyper@linkstart##1##2{}%
     \let\hyper@linkend\@empty\citep[#1][#2]{#3}}}
  \newcommandtwoopt{\citetads}[3][][]{\href{http://adsabs.harvard.edu/abs/#3}%
    {\def\hyper@linkstart##1##2{}%
     \let\hyper@linkend\@empty\citet[#1][#2]{#3}}}
  \newcommandtwoopt{\citeyearads}[3][][]%
    {\href{http://adsabs.harvard.edu/abs/#3}
    {\def\hyper@linkstart##1##2{}%
     \let\hyper@linkend\@empty\citeyear[#1][#2]{#3}}}
\makeatother

\begin{document}

\title{Interstellar magnetic cannon targeting the Galactic halo} 

\subtitle{A young bubble at the origin of the Ophiuchus and Lupus molecular complexes} 

\author{J.-F.~Robitaille\inst{\ref{inst1},\ref{inst2}} \and A.~M.~M.~Scaife\inst{\ref{inst1}} \and E.~Carretti\inst{\ref{inst3}} \and M.~Haverkorn\inst{\ref{inst5}} \and R.~M.~Crocker\inst{\ref{inst10}} \and M.~J.~Kesteven\inst{\ref{inst4}} \and S.~Poppi\inst{\ref{inst3}} \and L.~Staveley-Smith\inst{\ref{inst8},\ref{inst9}}}

\institute{Jodrell Bank Centre for Astrophysics, School of Physics and Astronomy,
        \\ The University of Manchester, Oxford Road, Manchester M13 9PL, UK\label{inst1}
        \and
        Univ. Grenoble Alpes, CNRS, IPAG, 38000 Grenoble, France,~\email{jean-francois.robitaille@univ-grenoble-alpes.fr}\label{inst2}
        \and
        INAF Istituto di Radioastronomia, Via Gobetti 101, 40129 Bologna, Italy\label{inst3}
        \and
        Department of Astrophysics/IMAPP, Radboud University, P.O. Box 9010, NL-6500 GL Nijmegen, The Netherlands\label{inst5}
        \and
        Research School of Astronomy and Astrophysics, Australian National University, Canberra, ACT, Australia\label{inst10}
        \and
        CSIRO Astronomy and Space Science, PO Box 76, Epping, NSW 1710, Australia\label{inst4}
        \and
        International Centre for Radio Astronomy Research, M468, University of Western Australia, Crawley WA 6009, Australia\label{inst8}
        \and
        ARC Centre of Excellence for All-sky Astrophysics (CAASTRO), M468, University of Western Australia, Crawley, WA 6009, Australia\label{inst9}
        } 

\date{Accepted for publication in A\&A on June 27, 2018}

\abstract{We report the detection of a new Galactic bubble at the interface between the halo and the Galactic disc. We suggest that the nearby Lupus complex and parts of the Ophiuchus complex constitute the denser parts of the structure. This young bubble, $\lesssim 3$ Myr old, could be the remnant of a supernova and expands inside a larger HI loop that has been created by the outflows of the Upper Scorpius OB association. An HI cavity filled with hot X-ray gas is associated with the structure, which is consistent with the Galactic chimney scenario. The X-ray emission extends beyond the west and north-west edges of the bubble, suggesting that hot gas outflows are breaching the cavity, possibly through the fragmented Lupus complex. Analyses of the polarised radio synchrotron and of the polarised dust emission of the region suggest the connection of the Galactic centre spur with the young Galactic bubble. A distribution of HI clumps that spatially corresponds well to the cavity boundaries was found at $V_{LSR} \simeq-100$ km s$^{-1}$. Some of these HI clumps are forming jets, which may arise from the fragmented part of the bubble. We suggest that these clumps might be `dripping' cold clouds from the shell walls inside the cavity that is filled with hot ionised gas. It is possible that some of these clumps are magnetised and were then accelerated by the compressed magnetic field at the edge of the cavity. Such a mechanism would challenge the Galactic accretion and fountain model, where high-velocity clouds are considered to be formed at high Galactic latitude from hot gas flows from the Galactic plane.}

\maketitle 

\section{Introduction}

The Galactic centre is one of the most complex regions of the Galaxy. Its complicated star formation history and the presence of multiple massive star clusters have shaped the large-scale structures, which appear as a superposition of loops observed through the atomic hydrogen (HI) line emission, gamma-rays (the `Fermi bubbles'), and through synchrotron radio emission \citepads{ 1984ApJS...55..585H, 1995A&A...294L..25E, 2007ApJ...664..349W, 2010ApJ...724.1044S, 2015MNRAS.452..656V, 2015ApJ...808..107C}. \citetads{2015MNRAS.452..656V} identified four possible explanations for the origin of these loops: old and nearby supernova remnants, outflows from the Galactic centre, bubbles powered by OB associations, and magnetic field loops illuminated by synchrotron emission.

In this paper, we concentrate on the central part of the Galaxy that is dominated by Loop I, also called the North Polar Spur (NPS), the bright X-ray emission \citepads{1995A&A...294L..25E}, and large HI loops \citepads{1992A&A...262..258D}. \citetads{2007ApJ...664..349W} proposed a large-scale model to describe the two polarised synchrotron shells associated with the Loop I bright radio filament extending above $90^{\circ}$ in Galactic latitude. In this model, the Upper Scorpius (USco) OB subgroup played a recent role in the expansion of the Loop I bubble and is responsible for the X-ray emission observed toward the Loop. However, the author concedes that the proposed model is a simple approximation to a presumably more complex reality that probably requires many earlier stellar winds and supernova events to explain the features observed at smaller scales. A careful inspection of the HI data, X-ray, and polarised radio emission in this region has allowed us to identify a young Galactic bubble located at the interface between the halo and the Galactic plane that possesses all the characteristics that are typically attributed to the Galactic chimney model \citepads{1989ApJ...345..372N}. Our new model takes into account the configuration of the local magnetic field, the morphology of HI structures, and the formation of the two well-known molecular clouds with regions with ongoing star formation, the Ophiuchus and Lupus molecular complexes. The HI data also provide evidence that high-velocity clouds (HVCs) are probably associated with this region and interact with the bubble.

Our analysis is organised as follows: the data are presented in Section \ref{sec:data}, the region is described in detail in Section \ref{sec:description}, structures seen in polarisation are described in section \ref{sec:polarisation}, a detailed description of the HVCs present in this region is provided in section \ref{sec:HVCs}, and an outline of the model is presented in section \ref{sec:outline}. The discussion and conclusion are finally presented in sections \ref{sec:discussion} and  \ref{sec:conclusion}.

\section{Data\label{sec:data}}

This analysis is based on several large Galactic surveys. The gas column density ($N_{\textrm{H}}$) maps are derived from the Planck all-sky dust optical depth map at 353 GHz ($\tau_{353}$) \citepads{2014A&A...571A..11P}. This map was produced by fitting a modified black-body spectrum to Planck intensity observations at 353, 545, and 857 GHz and the IRAS observations at 100 $\mu$m. In order to estimate the $N_{\textrm{H}}$ map, we used the dust opacity found using Galactic extinction measurements of quasars, $\tau_{353}/N_{\textrm{H}}=1.2 \times 10^{-26}$ cm$^2$ \citepads{2014A&A...571A..11P, 2016A&A...586A.138P}.

The HI maps are taken from the Parkes Galactic All Sky Survey (GASS) \citepads{2009ApJS..181..398M, 2010A&A...521A..17K}. The survey has an effective angular resolution of 14.4 arcmin at a velocity resolution of 1.0 km s$^{-1}$ and covers all Milky Way velocities between $V_{LSR}$ $-400$ km s$^{-1}$ and 500 km s$^{-1}$. The X-ray maps are taken from the ROSAT all-sky survey R5 band (0.56--1.21 keV) \citepads{1997ApJ...485..125S}. The maps have a pixel size of $12\times12$ arcmin.

Finally, the radio polarisation data are from the S-band Polarization All Sky Survey (S-PASS; \citetads{2013Natur.493...66C}; Carretti et al., in preparation). S-PASS is a single-dish polarimetric survey of the entire southern sky at 2.3 GHz, performed with the Parkes 64 m Radio Telescope, and its S-band receiver has a beam width FWHM = 8.9 arcmin. Final maps were convolved with a Gaussian window of FWHM = 6 arcmin to a final resolution of 10.75 arcmin.

\section{Description of the region\label{sec:description}}

\subsection{Molecular clouds and HI shells\label{sec:region}}

\begin{figure*}
\centering
\includegraphics[width=0.77\textwidth]{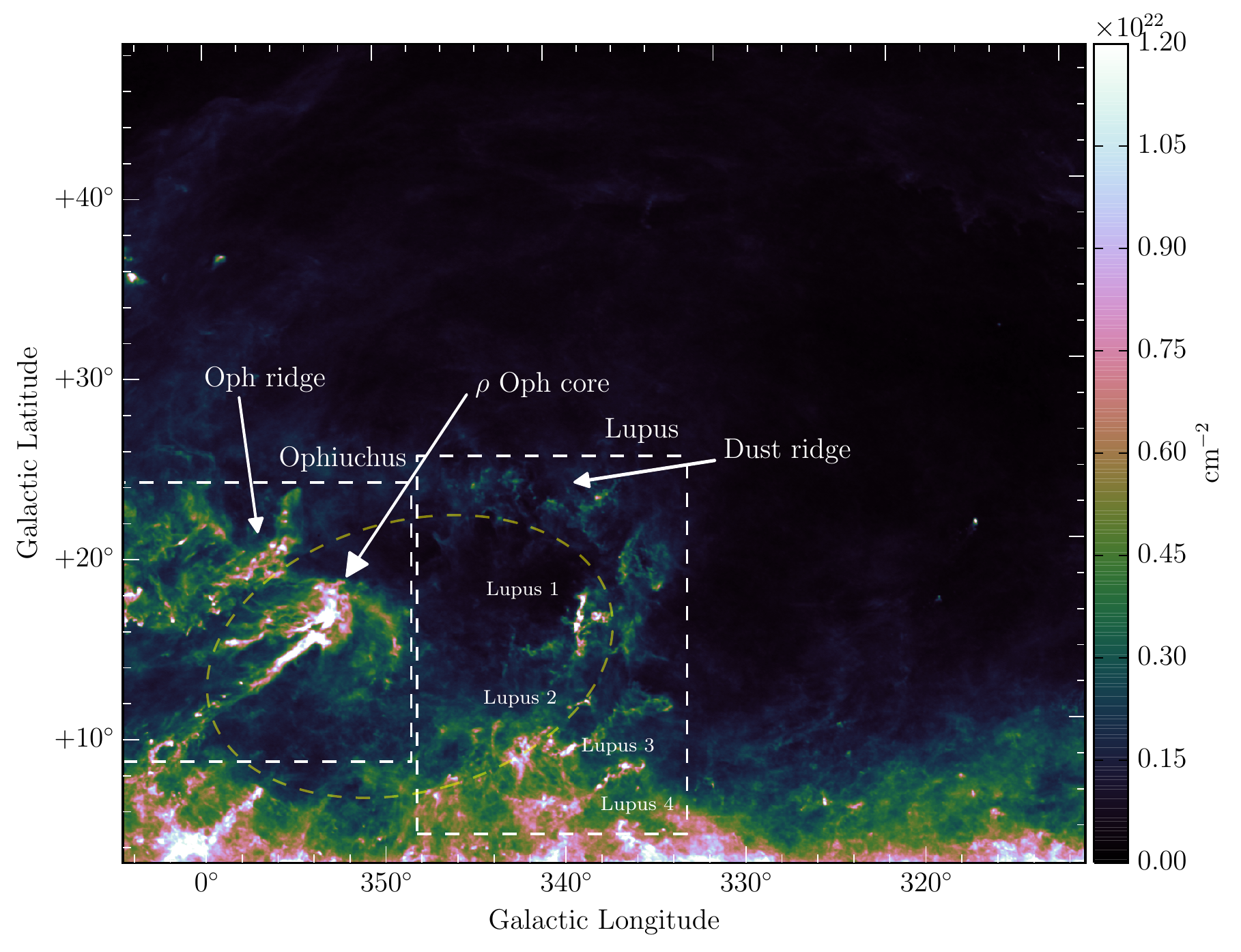}
\caption{Total gas column density ($N_{\textrm{H}}$) map derived from the $\tau_{353}$ Planck map. The white dashed rectangles delimitthe  Ophiuchus and Lupus dark complexes according to \citetads{2008A&A...489..143L}. The yellow dashed ellipse delimits the inner boundary of the bubble.}
\label{fig:dust850}
\end{figure*}

\begin{figure*}
\centering
\includegraphics[width=0.77\textwidth]{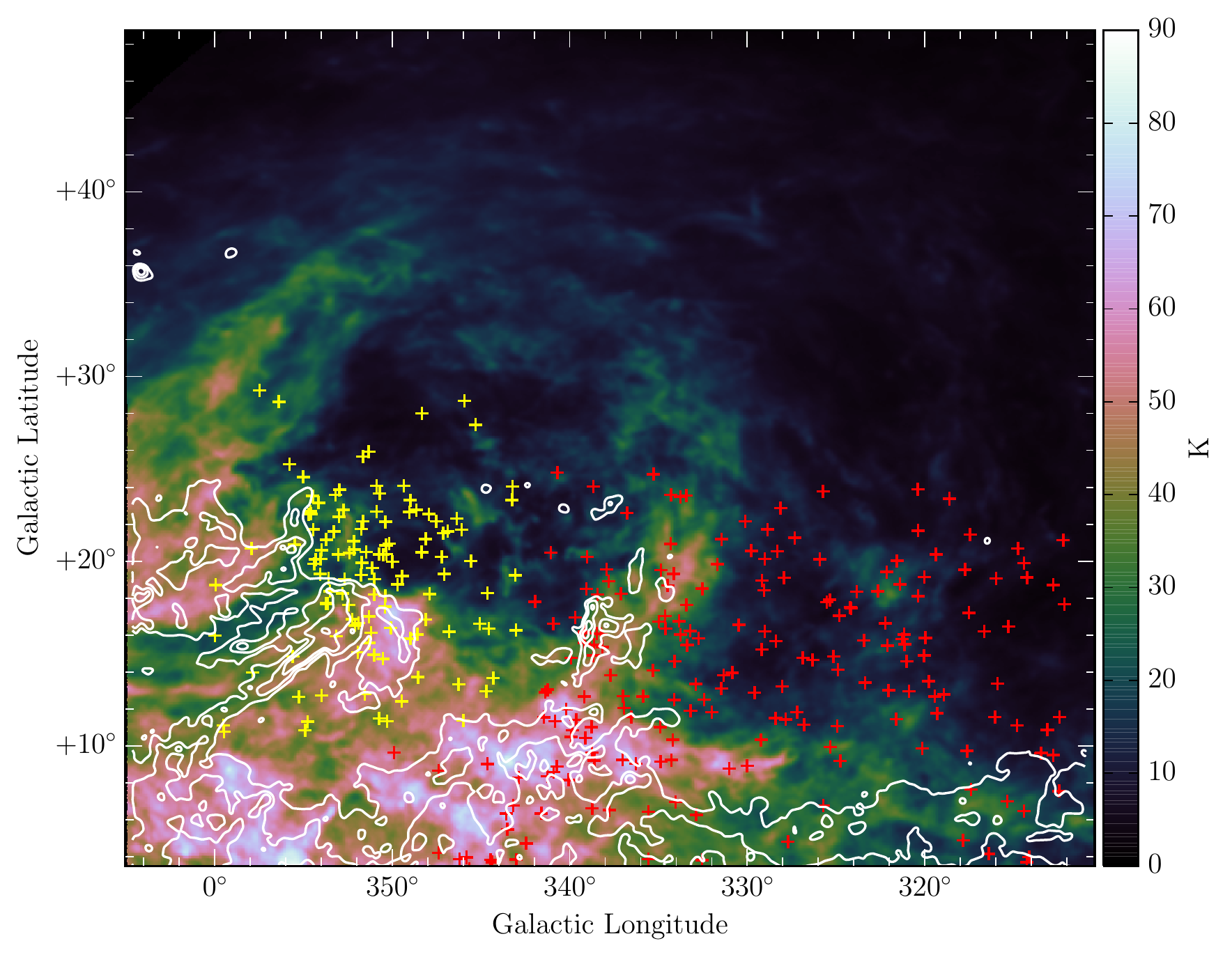}
\caption{USco HI shell at $V_{LSR}=5.77$ km s$^{-1}$ from GASS HI data. The white contours represent the $N_{\textrm{H}}$ map smoothed with a 5-pixel $\sigma$ Gaussian kernel, where the contour levels are at 2.5, 4.5, and 8.0$\times 10^{21}$ cm$^{-2}$. The yellow and red crosses show the position of the USco and UCL subgroups, respectively, according to the HIPPARCOS catalogue \citepads{1999AJ....117..354D}.}
\label{fig:HI_dust_HIP}
\end{figure*}

The region is well known for its two nearby molecular clouds, the Ophiuchus and the Lupus dark complexes. Figure \ref{fig:dust850} shows the Planck gas column density ($N_{\textrm{H}}$) map including the two molecular cloud complexes. The yellow dashed ellipse delimits the inner boundaries of the bubble that are discussed in this paper. Other important features for our analysis are pointed out by arrows. \citetads{2008A&A...480..785L} assigned distances of ($119 \pm 6$) pc and ($155 \pm 8$) pc to the $\rho$ Ophiuchus and the Lupus complex, respectively, according to HIPPARCOS distance estimates. Both complexes are part of the Gould Belt, a ring of molecular clouds and OB associations in the Galactic Plane, and have recently been analysed by \citetads{2015MNRAS.450.1094P} and \citetads{2017MNRAS.467..812M} using the JCMT Gould Belt Survey \citepads{2007PASP..119..855W} and the \textit{Herschel} Gould Belt Survey \citepads{2010A&A...518L.102A}. They are considered to be the regions of star formation closest to the Sun and are probably strongly affected by the nearby Scorpius-Centaurus OB association. \citetads{1992A&A...262..258D} showed that the Lupus complex is embedded in an expanding HI shell associated with the USco OB subgroup. Figure \ref{fig:HI_dust_HIP} shows the large HI shell at $V_{LSR}=5.77$ km s$^{-1}$ that extends over $b \sim 30^{\circ}$ extracted from GASS. The position of the USco and Upper-Centaurus Lupus (UCL) subgroups are also marked in Fig. \ref{fig:HI_dust_HIP}. An HI counterpart of the $\rho$ Ophiuchus core is visible inside the shell. For both molecular cloud complexes, Ophiuchus and Lupus, most of the HI is associated with the low total gas column density ($N_{\textrm{H}} \lesssim 2.5 \times 10^{21}$).

\citetads{2015A&A...584A..36G} (hereafter G15) proposed in their study that the large-scale compression due to the expanding USco HI shell and the UCL OB association wind located on the far side of the Lupus 1 molecular cloud would cool atomic gas sufficiently for the formation of molecules and for triggering star formation inside the Lupus 1 molecular cloud. The hypothesis that Lupus 1 has recently been shocked by expanding shells agrees with recent SCUBA-2 observations that were analysed by \citetads{2017MNRAS.467..812M}, who found a high number of Class 0/I young stellar object (YSO) sources and a lower number of Class III YSOs. This high ratio (1:1) suggests that Lupus 1 is young.

\begin{figure*}
\centering
\includegraphics[width=0.77\textwidth]{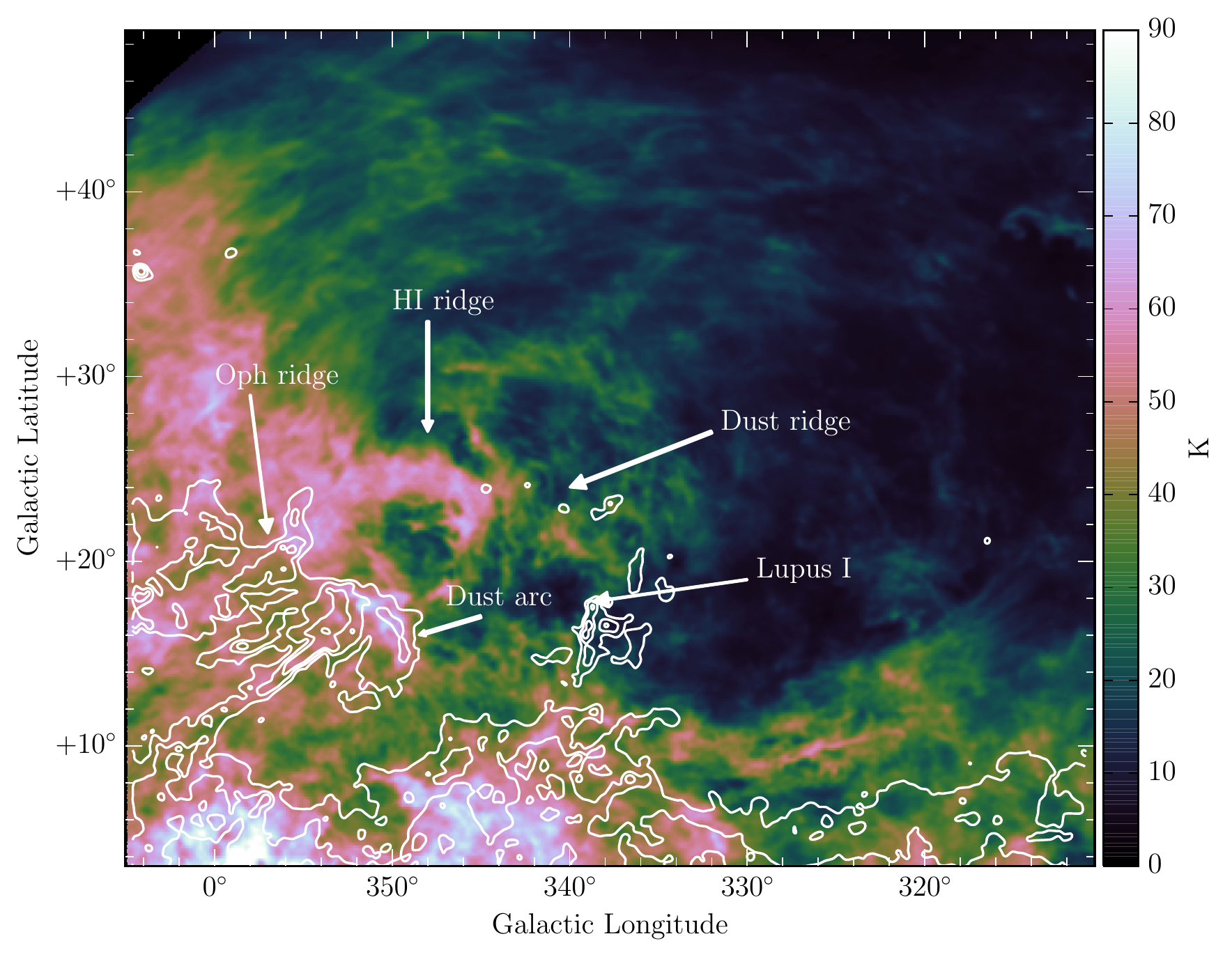}
\caption{HI ridge at $V_{LSR}=1.65$ km s$^{-1}$ from GASS HI data. The white contours represent the $N_{\textrm{H}}$ map in the same way as in Fig. \ref{fig:HI_dust_HIP}.}
\label{fig:HI_dust_bubble}
\end{figure*}

G15 also reported two dust voids located on the eastern and western side of the Lupus 1 molecular cloud. The eastern void is delimited by a ring-like dust ridge that contains several molecular clouds located to the north of Lupus 1 (see the dust ridge in Fig. \ref{fig:dust850}). The authors explain these two voids by the cumulative feedback from massive stars in the USco and UCL subgroups, which would have triggered the star formation in the region. We suggest a new model to explain the morphology of this region. In this model, the eastern void is the location of a nearby Galactic bubble that is probably expanding inside the older USco HI shell. We also suggest that the western void, rather than being produced by the UCL subgroups feedback, is probably filled with hot gas escaping from the bubble through the fragmented Lupus molecular cloud.

\begin{figure}
\centering
\includegraphics[width=0.45\textwidth]{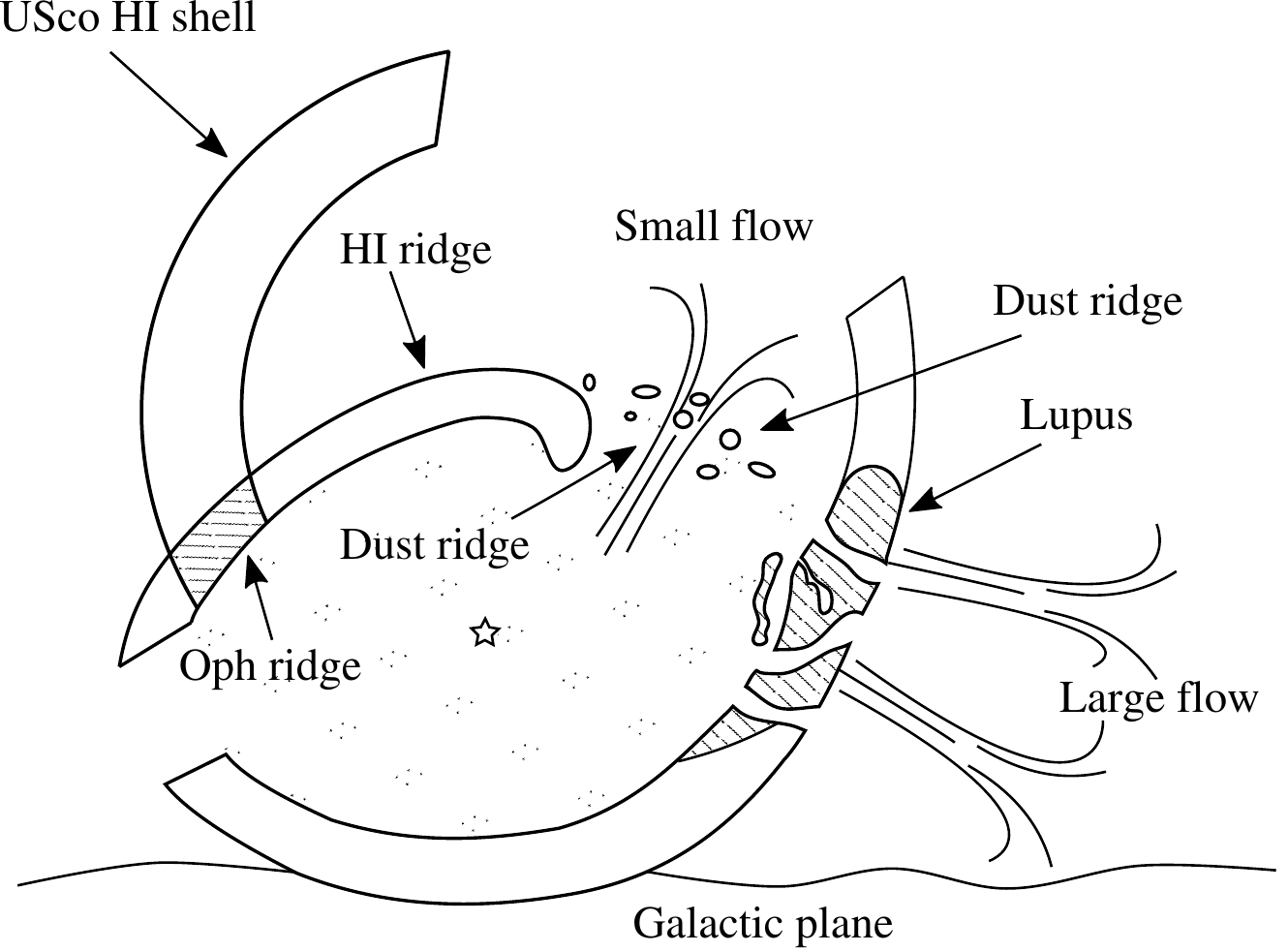}
\caption{Schematic diagram of the bubble at the interface between the Galactic disc and the halo seen from the same point of view as the observations. The smaller bubble filled with hot X-ray gas is filled with black dots. A star marks the centre of the bubble. Parallel lines mark the location of the Lupus molecular cloud and Oph ridge where the bubble meets the USco HI shell.}
\label{fig:Model}
\end{figure}

Figure \ref{fig:HI_dust_bubble} shows the HI at local standard of rest (LSR) velocity of $1.65$ km s$^{-1}$. The contours indicate the position of the molecular clouds according to the $N_H$ map. It is important to note the presence of a complementary HI ridge completing the ring-like dust ridge described by G15. The lower column density gas associated with the HI ridge is also visible in the $N_{\textrm{H}}$ map of Fig. \ref{fig:dust850}. In addition, on the left-hand side of the HI ridge, another ridge-like structure can be identified. This structure is part of the Ophiuchus north molecular clouds and has been labelled the Oph ridge in this model. The $^{13}$CO cloud associated with the Oph ridge is located at $V_{LSR}=0$--$2$ km s$^{-1}$ \citepads{1991ApJS...77..647N}, which corresponds well to the velocity of the HI ridge. An HI counterpart of the Oph ridge is also visible around 5.77 km s$^{-1}$ (see Fig. \ref{fig:HI_dust_HIP}). In our model, the HI ridge, which appears to complete the upper boundary of the eastern void with the dust and Oph ridges, is part of a fragmented bubble structure with the Lupus molecular complex. A schematic diagram of the bubble embedded in the USco HI shell is shown in Fig. \ref{fig:Model}.

The distance of $155 \pm 8$ pc for the Lupus was assigned by \citetads{2008A&A...480..785L} using the whole molecular complex. However, with recent parallax measurements, \citetads{2013A&A...558A..77G} found significant depth effects in the molecular cloud. They measured distances of $182^{+7}_{-6}$, $167^{+19}_{-15}$, $185^{+11}_{-10}$ and $204^{+18}_{-15}$ pc for the brightest subclouds Lupus 1, 2, 3, and 4 (see labels in Fig. \ref{fig:dust850}). They classified most of the young stars located in the clouds following the outer boundaries of the ellipsoid shape in Fig. \ref{fig:dust850} as `off-cloud' stars with a measured distance of $139^{+10}_{-9}$ pc. This distance is consistent with the estimated distance of 130 pc for the Oph ridge \citepads{2012ApJ...754..104H}. Consequently, the distance of $139^{+10}_{-9}$ pc is assigned to the bubble model in this paper. The hypothesis that Lupus 1 has recently been  shocked by an expanding shell, as claimed by G15 and \citetads{2017MNRAS.467..812M}, is still consistent with our interpretation of the region's history. The inhomogeneity of the material before the bubble expansion can explain the fragmented morphology of the northwestern side of the bubble. The multiple compression scenario, where the bubble expansion colliding with the larger USco HI shell would be at the origin of the Lupus molecular complex, will also be explored in the discussion section.

This model places the Lupus complex at the edge of a shell-like structure that is probably filled with hot X-ray gas (see next section). We propose that its fragmented morphology might also be at the origin of the outflow of hot gas seen in the western void.

\subsection{Galactic chimney model: X-ray emission and outflows \label{sec:chimney}}

\begin{figure*}
\centering
\includegraphics[width=0.77\textwidth]{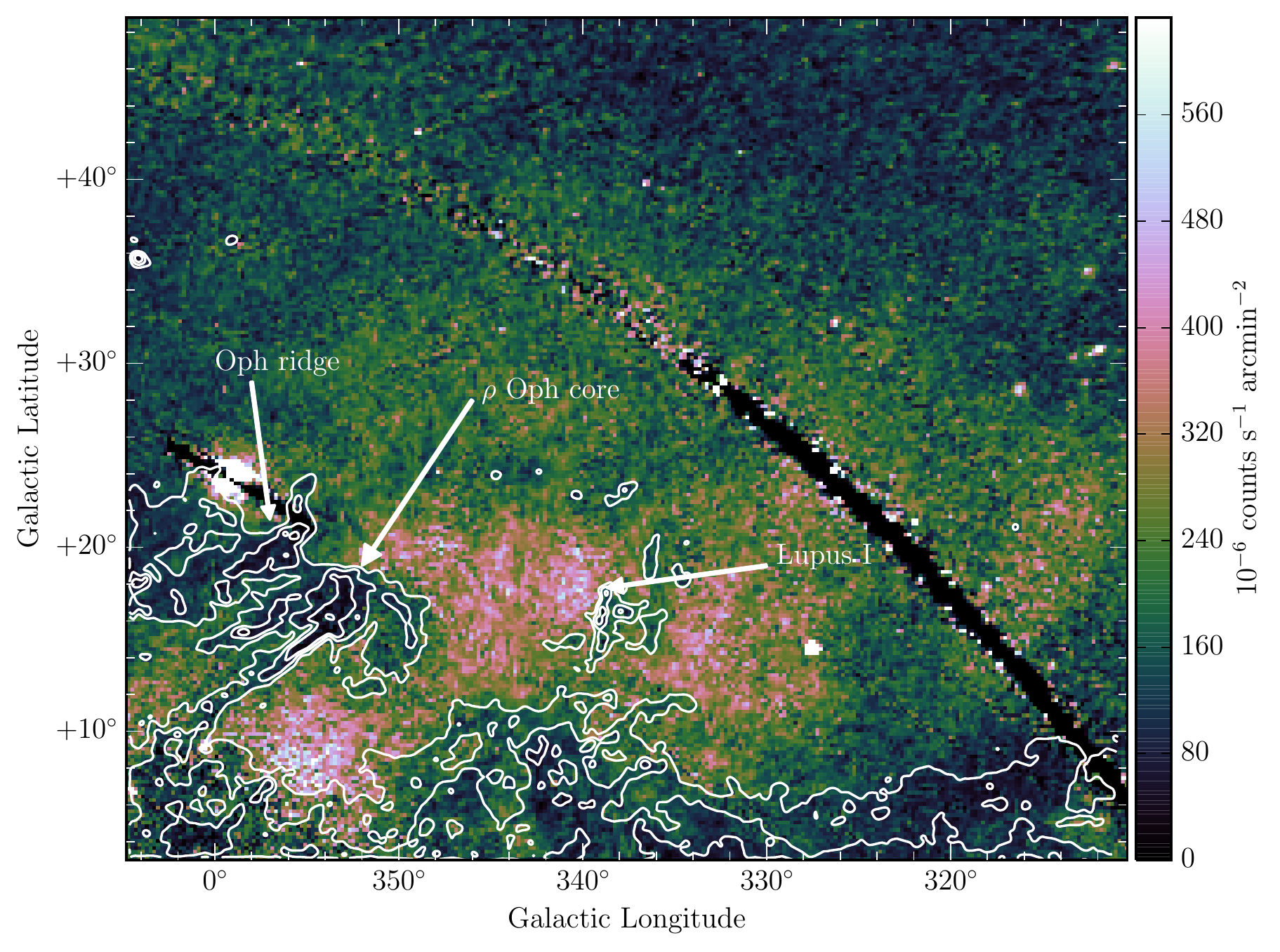}
\caption{Colours show the X-ray emission (ROSAT R5 band) for the same region as shown in Figs. \ref{fig:dust850}, \ref{fig:HI_dust_HIP}, and \ref{fig:HI_dust_bubble}. The white contours represent the $N_{\textrm{H}}$ map in the same way as in Fig. \ref{fig:HI_dust_HIP}.}
\label{fig:X-ray_NH_bubble}
\end{figure*}

As described by G15, the two dust voids on either side of Lupus 1 are filled with hot X-ray gas. Figure \ref{fig:X-ray_NH_bubble} shows the X-ray emission (ROSAT R5 band) for the same region as shown in the previous figures. G15 noted that since the X-ray gas follows the edge of Lupus 1, it could indicate that the molecular cloud is embedded in the hot gas. However, they did not rule out that Lupus 1 might also be seen in absorption against the hot gas emission, as is also probably the case for the Ophiuchus cloud complex. 

\begin{figure*}
\centering
\includegraphics[width=0.77\textwidth]{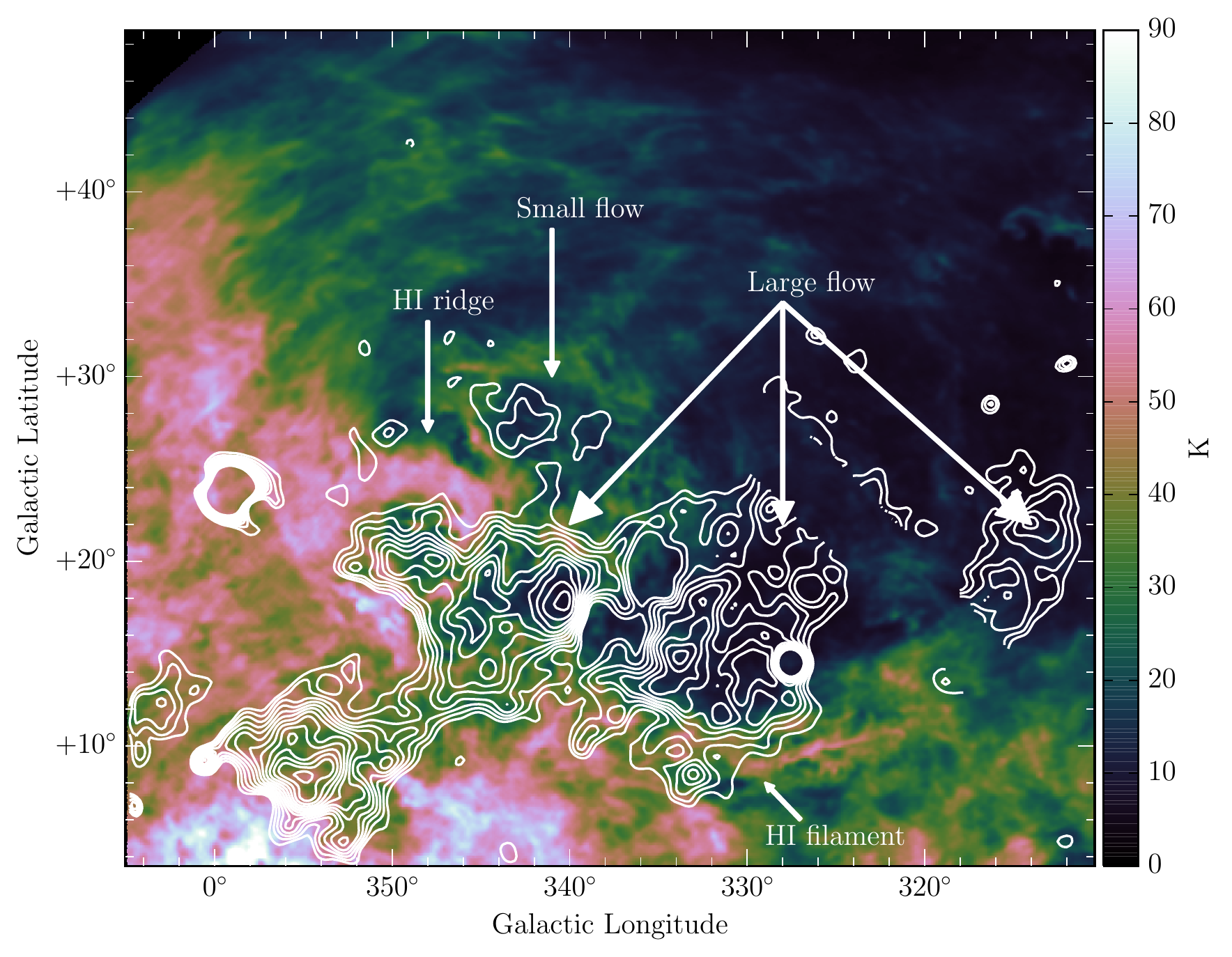}
\caption{HI emission at $V_{LSR}=1.65$ km s$^{-1}$ overlaid with the X-ray map smoothed with a 2-pixel $\sigma$ Gaussian kernel, where contours start from 225 to $500\times10^{-6}$ counts s$^{-1}$ arcmin$^{-2}$ with an interval of $20\times10^{-6}$ counts s$^{-1}$ arcmin$^{-2}$. The contours are masked between longitude 225.5$^{\circ}$ and 229$^{\circ}$ in ecliptic coordinates, because of the stripe seen in Fig. \ref{fig:X-ray_NH_bubble}.}
\label{fig:HI_Xray_bubble}
\end{figure*}

Figure \ref{fig:HI_Xray_bubble} shows the HI emission at $V_{LSR}=1.65$ km s$^{-1}$ shown in Fig. \ref{fig:HI_dust_bubble} overlaid with the X-ray contours. At this velocity, the HI gas seems to encompass the X-ray emission. This connection between the X-ray and the HI at 1.65 km s$^{-1}$ seems to continue on the right-hand side of the Lupus molecular complex as well, where a sharp HI filament at a Galactic latitude of $b\sim10^{\circ}$ corresponds well to the southern X-ray boundary. 

In the west-northwest part of the HI cavity shown in Fig. \ref{fig:HI_Xray_bubble}, the HI ridge and the southern HI filament are separated by a breach. If the X-ray emission is considered as continuous behind the Lupus complex, or through the Lupus complex, then the hot gas filling this region can be associated with an outflow of gas from the HI cavity. This outflow can be divided into a small flow, located above the dust ridge on the right-hand side of the HI ridge, and a large flow, located below the dust ridge, which would pass behind, or through, the Lupus molecular cloud. The large flow is subdivided into in three parts, one filling the HI cavity and located on the left-hand side of Lupus, one located on the right-hand side of Lupus, and the third located at the edge of the field around $l\sim315^{\circ}$. The gap between the middle part of the flow and western part could be explained by a shadow effect caused by colder gas located in front of the flow shown in HI emission at 1.65 and 5.77 km s$^{-1}$ in Figs. \ref{fig:HI_dust_HIP} and \ref{fig:HI_dust_bubble}. It is interesting to note that the small flow seen in X-rays is also embedded in what seems to be an HI flow located above the dust ridge. 

Again, as for the Ophiuchus and the Lupus complex, the HI ridge could also produce a shadow effect over the hot gas. However, unlike the western side of the Lupus, the northern side of the HI ridge has a significantly lower X-ray brightness than the southern part inside the cavity. This brightness variation is consistent with hot gas enclosed by colder HI gas. Furthermore, the total column density map in Fig. \ref{fig:dust850} does not indicate any amount of matter along the line of sight that would be sufficient to produce a shadow effect on the northern side of the HI ridge. According to these observations, we conclude that it is highly probable that the X-ray emission on the eastern side of Lupus is emitted by local hot gas contained inside the bubble. This morphology corresponds well to the Galactic chimney scenario located at the interface between the Galactic disc and the Galactic halo, as proposed by \citetads{1989ApJ...345..372N}. This interpretation also agrees with the analysis of \citetads{2014A&A...566A..13P}, which compares the X-ray bright regions of the R5 band ROSAT map with a three-dimensional interstellar medium (ISM) map computed from extinction measurements \citepads{2014A&A...561A..91L}. They found that the three bright X-ray regions located at $l,b=$(330$^{\circ}$,14$^{\circ}$), (342$^{\circ}$,18$^{\circ}$) and (355$^{\circ}$,9$^{\circ}$) correspond well to a nearby cavity in the local ISM distribution that is located at a distance of $\sim 130$ pc. This distance corresponds to the distance suggested for the bubble.

\section{Polarisation data \label{sec:polarisation}}

\begin{figure*}
\centering
\includegraphics[width=1.0\textwidth]{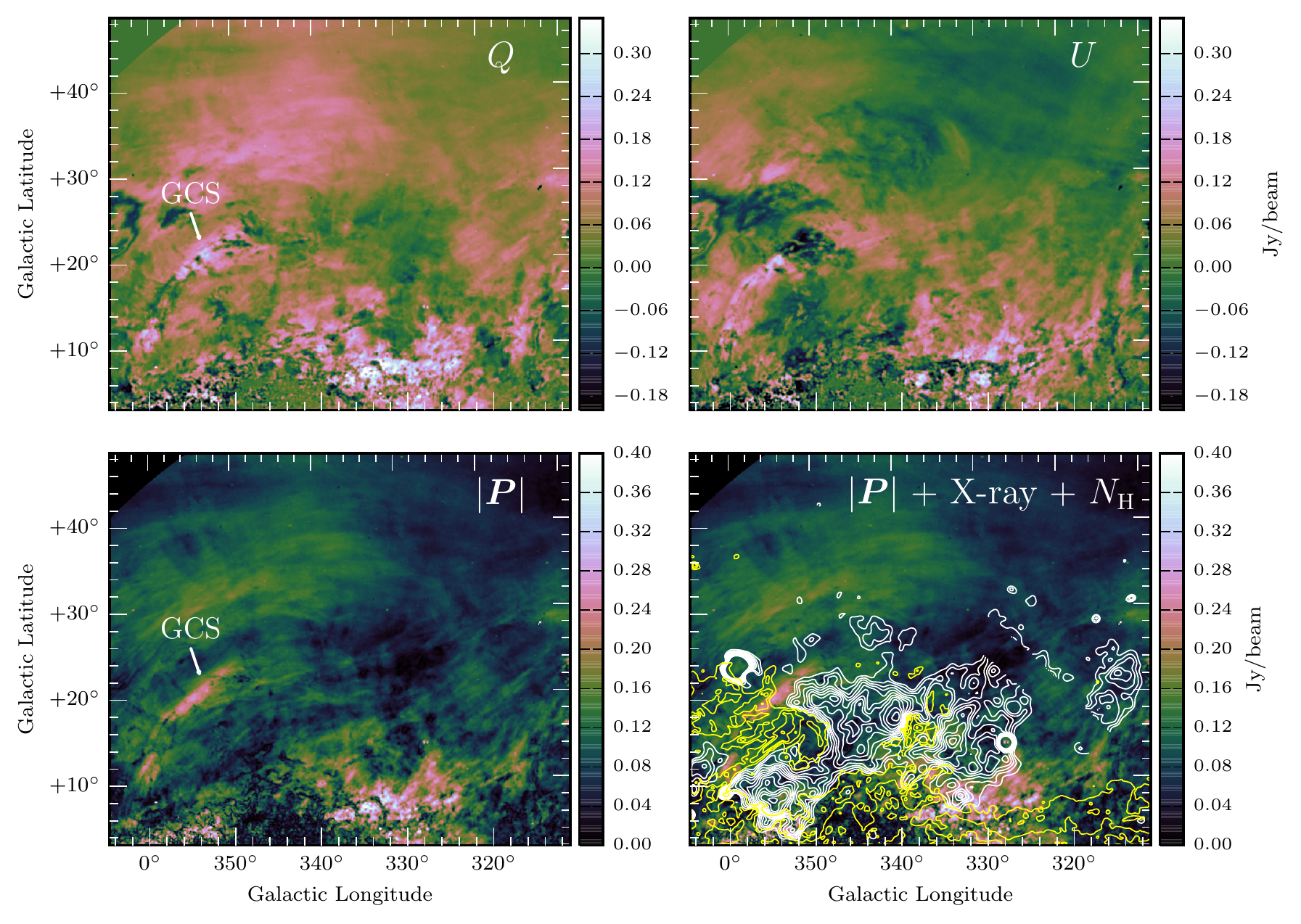}
\caption{Polarisation maps from S-PASS at 2.3 GHZ. Top panels: Stokes $Q$ (left) and $U$ (right) parameters. Bottom panels: Polarisation intensity on the left and on the right of the polarisation intensity overlaid with the X-ray intensity (white) and the total column density (yellow).}
\label{fig:QUP_bubble}
\end{figure*}

This section presents the analysis of the 2.3 GHz S-PASS polarisation data associated with this region. The Stokes $Q$ and $U$ maps are shown in the top panels of Fig. \ref{fig:QUP_bubble}. The main structures visible in Stokes $Q$ and $U$ on the eastern part of the field are the two HII regions Sh 2-7 and Sh 2-27, and the arc of the Galactic Centre Spur (GCS), which extends from $b\sim3^{\circ}$ to $\sim24^{\circ}$. The left lower panel in Fig. \ref{fig:QUP_bubble} shows the polarisation intensity map of the field, $|\boldsymbol{P}|=\sqrt{Q^2+U^2}$. Around $l,b \simeq (3^{\circ}, 21^{\circ})$, some depolarisation canals associated with sharp changes of the polarisation angles at spatial scales lower than the beam can be observed. These depolarisation canals are related to the HII region Sh 2-27 located around the runaway star $\zeta$ Oph. The two HII regions are located at a distance close to the assumed distance of the bubble, $\sim 140$ pc. Since both HII regions are visible in $Q$ and $U$, and through the polarisation gradient \citepads{2014A&A...566A...5I, 2017MNRAS.468.2957R}, and not in polarisation intensity except for the depolarisation canals, we can conclude that structures observed in $Q$ and $U$ are mainly caused by Faraday rotation. The possible link between $\zeta$ Oph and the USco HI shell has first been explored by \citetads{2000ApJ...544L.133H} and is analysed in more detail in section \ref{sec:origin}.

\citetads{1989ApJ...341L..47S} proposed two interpretations for the location of the GCS: a local shell-like object produced by an old supernova remnant, or an ejection feature associated with Galactic centre activity. The authors concluded that the latter was more probable since the spur lacks sharp edges and no HI structure seems to be associated with the spur. \citetads{2013Natur.493...66C} also pointed out the depolarisation at its base is due to the ISM in the Sagittarius arm, setting a lower limit for the distance of this section of the GCS to some 2 kpc. The GCS association with the bubble described in this paper is uncertain. However, the good spatial correspondence between the eastern edge of the bubble and the GCS, from $b\sim 10^{\circ}$ and above, justifies a new inspection of the structure.

\begin{figure*}
\centering
\includegraphics[width=1.0\textwidth]{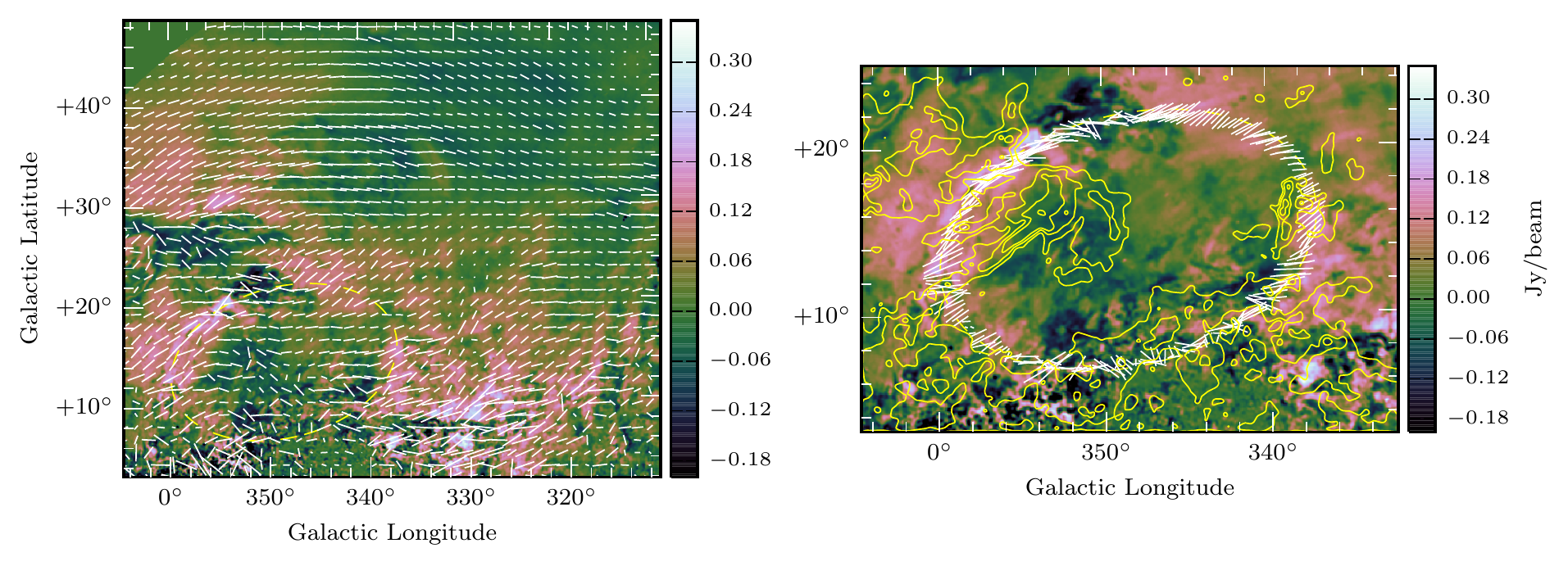}
\caption{Stokes $U$ map overlaid with polarisation vectors. The vector direction shows the orientation of the magnetic field, and the length is proportional to the polarisation intensity. The right panel shows the polarisation vectors only for an ellipse with a thickness of 1$^{\circ}$ , which defines the bubble boundaries. The outer limit of the ellipse is delimited by the yellow dashed line in both panels. The ellipse is centred on $l=348.5^{\circ}$ and $b=14.5^{\circ}$, it has a major axis of $23^{\circ}$, a minor axis of $15^{\circ}$ , and a tilt angle of $15.5^{\circ}$. The yellow contours show the total gas column density $N_H$ , where the contour levels are at 2.5, 4.5, and 8.0$\times 10^{21}$ cm$^{-2}$.\label{fig:Bubble_polarisation}}
\end{figure*}

In addition to its spatial correspondence with the eastern edge of the bubble, the GCS possess unique properties that are very different from the other parts of the bubble. First of all, the GCS possess a stronger polarisation intensity than the other edge sections of the bubble. Secondly, as seen in Fig. \ref{fig:Bubble_polarisation}, the magnetic field orientation is mostly perpendicular to the edge of the cavity everywhere, and the magnetic field orientation associated with the GCS is parallel to the spur. Recent simulations of a supershell by \citetads{2017A&A...599A..94N} show that with time, due to the strong compression, the projected magnetic field tends to be oriented along the collision surface and its strength becomes significantly stronger than the field in the diffuse hot medium. The orientation of the magnetic field on the eastern and northeastern part of the cavity corresponds well to this description. In this area, the material swept up by the shell expansion could have compressed the magnetic field lines, which appear to be oriented along the collision surface. If the GCS is associated with the cavity, the compressed magnetic field could possibly be at its origin, which would also explain the stronger polarisation intensity. 

A closer look at the right panel of Fig. \ref{fig:Bubble_polarisation} allows us to establish where the polarisation angle changes around the cavity. In this figure, the magnetic field orientation is shown only for an ellipse with a thickness of 1$^{\circ}$ at the boundaries of the cavity. The vector are plotted over the Stokes $U$ map. Polarisation angles start to be parallel to the GCS at $b\sim14^{\circ}$ and change direction again at $b\sim21^{\circ}$. Between $(l,b)\sim(354^{\circ}, 21^{\circ})$ and $\sim(348^{\circ}, 22^{\circ}), $ the vector angles vary significantly to become perpendicular on the western edge until the Galactic plane causes depolarisation at the bottom of the cavity.

It is interesting to note that the section of the spur in which the magnetic field orientation is parallel corresponds well to the location of the inner part of the Oph ridge (see contours in the right panel of Fig. \ref{fig:Bubble_polarisation} and the colour scale of Fig. \ref{fig:dust850}). According to the simulations of \citetads{2017A&A...599A..94N} and to the morphology of the region, the Oph ridge might represent the material that is swept up by the shell expansion that caused the cavity. The same shell expansion would also be responsible for the compressed and enhanced magnetised medium traced by the GCS. On the other hand, the last section of the spur between $(l,b)\sim(354^{\circ},
21^{\circ})$ and $\sim(348^{\circ}, 22^{\circ})$, with its almost random polarisation angles and negative Stokes $U,$  corresponds well to the central position below the HI ridge before the hook (see Fig. \ref{fig:HI_dust_bubble} and \ref{fig:HI_Xray_bubble}). This particular section of the shell shows a transition from dense molecular gas to atomic gas, that is, from the Oph ridge to the HI ridge. High anisotropies in the initial density of the region might explain the difference in density and the difference in the strength and direction of the magnetic field that is located close to the inner part of the shell. Another possibility is that the polarisation orientation for this section of the GCS is affected by the HII region Sh2-7 acting as a Faraday screen in front of the spur.

Again, according to the results of \citetads{2017A&A...599A..94N}, if less dense gas is present on the west side of the expanding bubble, the magnetic field would have maintained its initial orientation along the direction of the expansion. This characteristic seems to be true for the transition section between the Oph ridge and the HI ridge, and also for the western and north-western part of the bubble. Compared to the Oph ridge on the eastern side, the Lupus molecular cloud has a large-scale magnetic field oriented perpendicularly to its filamentary morphology. This characteristic has also been noted by BLASTPol observations \citep{2014ApJ...784..116M} and by \citet{2015MNRAS.453.2036B}, who showed that many perpendicular branches connected to the Lupus filaments are aligned with the magnetic field. For this side of the bubble, possible anisotropies in the local initial density and a lesser amount of gas than at the eastern side could explain the fragmented morphology of the Lupus molecular complex and the orientation of the magnetic field perpendicular to the western edge of the cavity. As described in section \ref{sec:chimney}, the fragmented morphology of the Lupus complex might also have allowed the hot gas inside the bubble to escape through the eastern part of the bubble.

Furthermore, it is important to note that the Lupus molecular cloud is located precisely where the bubble described in our model overlaps with the larger USco HI shell described by \citetads{1992A&A...262..258D}. This property can also be attributed to the Oph ridge, although the orientation of the ridge does not match the orientation of the larger shell (see the schematic diagram in Fig. \ref{fig:Model}). This particular location of the molecular clouds as well as the orientation of the magnetic field inside the bubble correspond well with the model of molecular cloud formation of \citetads{2015A&A...580A..49I} which is based on the results of recent high-resolution magnetohydrodynamical simulations. In their model, molecular clouds are formed in limited regions where dense HI shells driven by expanding bubbles overlap. The formation of molecular clouds needs multiple compressions where the compressional direction is nearly parallel to the local mean magnetic field lines. This model corresponds well with our analysis of the Lupus complex environment, where the magnetic field is parallel to the expansion direction of the bubble and the bubble location overlaps the USco HI shell. In the case of the GCS, assuming that it is associated with the bubble, the gas compression leading to the formation of the Oph ridge has also compressed the magnetic field  and enhanced the field strength.

In that context, the morphology of the bubble shell is perhaps not only dependent on the initial density anisotropies of the region, but also on the general orientation of the magnetic field before the gas compression. According to the left panel of Fig. \ref{fig:Bubble_polarisation}, the general magnetic field direction of the region is parallel to the Galactic plane. This general orientation of the magnetic field could have facilitated the bubble expansion in the eastern and western directions, which could explain the ellipsoid shape of the cavity. At the northern edge of the cavity, where the magnetic field is nearly perpendicular to the bubble expansion front propagation, the bubble shell boundary is less dense and corresponds to the location of the HI ridge.

The comparisons mentioned above between the location of the GCS and the Oph ridge molecular cloud, together with the magnetic field orientation compared with the bubble compressional expansion might indicate that the GCS is a local structure associated with the bubble. However, some other findings indicate that the GCS is not associated with this local bubble. Carretti et al. (2013) found that the GCS base at $b\sim3^{\circ}$ is depolarised by ISM associated with the Sagittarius arm, setting its distance equal to or larger than 2 kpc. Moreover, the GCS below $b\sim10^{\circ}$ seems to depart from what is identified as the edge of the bubble. The lower left panel of Fig. \ref{fig:QUP_bubble} indeed shows a depolarisation of the GCS at a Galactic latitude lower than $b \sim 10^{\circ}$. On the other hand, the lower right panel reveals that the partial depolarisation of the GCS at $b \sim 10^{\circ}$, like the depolarisation between $b \sim 14^{\circ}$--$16^{\circ}$, also corresponds well with the position of the $\rho$ Ophiuchus core and its streamers (see section \ref{sec:Oph} for more details on the $\rho$ Oph core and its streamers). The currently available data are not sufficient to distinguish between these two interpretations, and more data are required.

The lower right panel in Fig. \ref{fig:QUP_bubble} also reveals an overlap of the X-ray emission with the GCS at $b \sim 21^{\circ}$. Free electrons in the hot X-ray gas can also generate polarisation angle rotation, which can contribute to the partial depolarisation of the GCS. The distance of the $\rho$ Oph core is estimated to be $137\pm1.2$pc \citepads{2017ApJ...834..141O}. This distance corresponds well to the assumed distance for the bubble. If the $\rho$ Oph core and its streamers are closely related to the bubble, as proposed in section \ref{sec:Oph}, then their position in front of the GCS would not necessarily reject the hypothesis that the GCS is a local structure associated with the bubble described in this paper. However, we recognise that the depolarisation of the GCS around $b \lesssim 10^{\circ}$ might also be attributed to ionised gas located much farther away, as far as 2.5 kpc, and in that case, it would identify the GCS as a non-local structure.

\section{High-velocity clouds\label{sec:HVCs}}

\begin{figure*}
\centering
\includegraphics[width=1.0\textwidth]{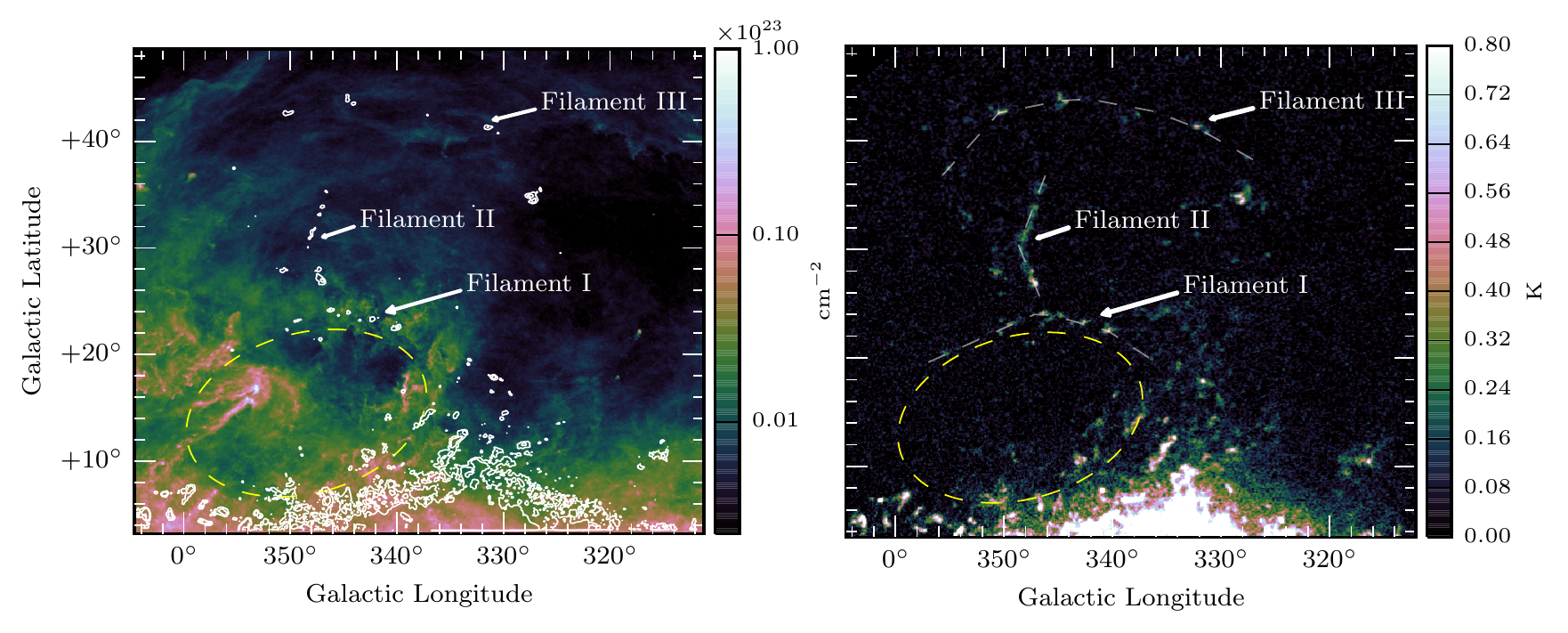}
\caption{Left panel: GASS data showing the HI emission at $V_{LSR}=-95.64$ km s$^{-1}$ with white contours. The contour levels are at 3.0 and 6.0 $K$. The background colour scale is the total gas column density $N_{\textrm{H}}$. The yellow dashed ellipse delimits the edges of the bubble as seen in Fig. \ref{fig:HI_dust_HIP}. Right panel: HI emission map at $V_{LSR}=-95.64$ km s$^{-1}$. The white dashed lines represent the slices used to produce the velocity-position plots in Fig. \ref{fig:pos-vel_HVC_b}.\label{fig:HVC_Planck}}
\end{figure*}

The region analysed in this paper is part of a larger region selected as a pilot field for GASS that was analysed by \citetads{2008ApJ...688..290F}. The analysis focused on HI halo clouds at negative velocities. At $V_{LSR}\simeq-95$ km s$^{-1}$ , they noted filamentary structures of clouds extending at high Galactic latitude. Figure \ref{fig:HVC_Planck} shows the HI emission at $V_{LSR}=-95.64$ km s$^{-1}$, where three filamentary structures made of HI clumps are visible. The lower filamentary structure (filament I) forms an arc located below the HI ridge at $V_{LSR}=1.65$ km s$^{-1}$ and the dust ridge. The second filamentary structure (filament II) is almost perpendicular to the first filament and extends from Galactic latitude $b\sim26^{\circ}$ to $\sim36^{\circ}$. The southern extremity of this filament is also located close to the interface between the HI ridge and the dust ridge. Finally, the third filament (filament III) forms a large arc located at a much higher Galactic latitude of $b\sim44^{\circ}$.

These clouds are located at a velocity $\sim100$ km s$^{-1}$ apart from the HI cavity associated with the bright X-ray emission. The spatial correspondence between HI structures seen at such different velocities is puzzling. According to \citetads{2008ApJ...688..290F}, if these HI clouds originate from the Galactic plane, for instance, HI pushed from the disc by violent supernovae and stellar winds or fragments of HI shells, and if they conserve their tangential velocity corresponding to Galactic rotation, their distances should correspond to $d_{t}=R_0\cos l/\cos b$, where $R_0\equiv8.5$ kpc is the radius of the solar circle. For a cloud located at $l\sim345^{\circ}$ and $b\sim25^{\circ}$ , the corresponding tangent distance is $\sim 9$ kpc. This is significantly more distant than the lower limit of $\sim130$ pc derived for the bubble according to the Lupus complex distance. These velocity and distance discrepancies suggest that if the HI clumps are associated with the HI cavity, at least in the case of filaments I and II, they have probably been accelerated by physical mechanisms other than Galactic rotation. Thus these particular HI clumps correspond well to the definition of high-velocity clouds (HVCs), that is, atomic HI cloud complexes located at high Galactic latitudes that move with high velocities that do not match a simple model of circular rotation of our Galaxy \citepads{1997ARA&A..35..217W}. The limit of HVCs is usually defined at $|V_{LSR}|\gtrsim90$ km s$^{-1}$.

\begin{figure}
\centering
\includegraphics[width=0.5\textwidth]{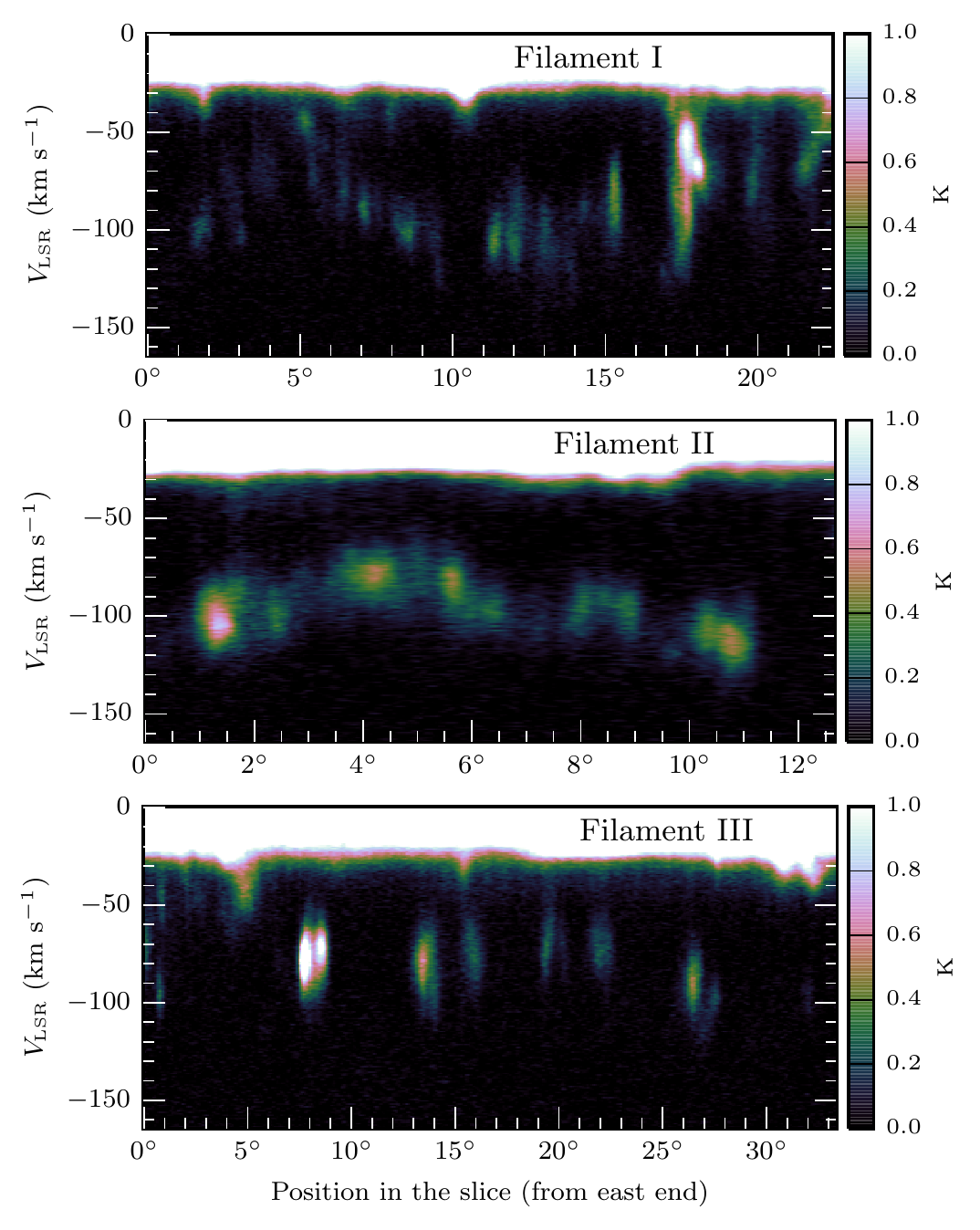}
\caption{Velocity-position plots for the three filaments shown in Fig. \ref{fig:HVC_Planck}.\label{fig:pos-vel_HVC_b}}
\end{figure}

The velocity-position plots for the three filaments are shown in Fig. \ref{fig:pos-vel_HVC_b}. The slices used to produce the velocity-position plots are indicated by the white dashed lines in the right panel of Fig. \ref{fig:HVC_Planck}. The width of the slices is $0.5$ degree, and they are centred on the position of the white dashed lines. For filament I, the velocity-position plot forms a kinematic arc between $\sim5^{\circ}$ and $18^{\circ}$ that connects with velocities corresponding to the Galactic plane at $V_{LSR}\gtrsim-30$ km s$^{-1}$. The kinematic arc might be evidence for an acceleration of the HI clouds inside the bubble and along filament I. The maximum velocity at the bottom of the arc is $\sim-105$ km s$^{-1}$ and corresponds well to the velocity of the first HI clump of filament II located on the left-hand side in the second panel of Fig. \ref{fig:pos-vel_HVC_b}. The velocity of filament II is more constant, and assuming that all clumps of filament II have the same origin, the opposite curvature of the filament compared to filament I might suggest a deceleration of some clumps. As reported by \citetads{2008ApJ...679L..21L} (see also \citetads{2013ApJ...777...55H}) in the analysis of the velocity-position slices for the Smith Cloud, such deceleration can suggest a clump interaction with the lower density ISM at higher Galactic latitude or simply that the slower clumps acquire less acceleration before escaping the bubble. Filament III has a velocity-position plot similar to filament II, but at a lower velocity of $V_{LSR}\sim-75$ km s$^{-1}$.

\begin{figure*}
\centering
\includegraphics[width=1.0\textwidth]{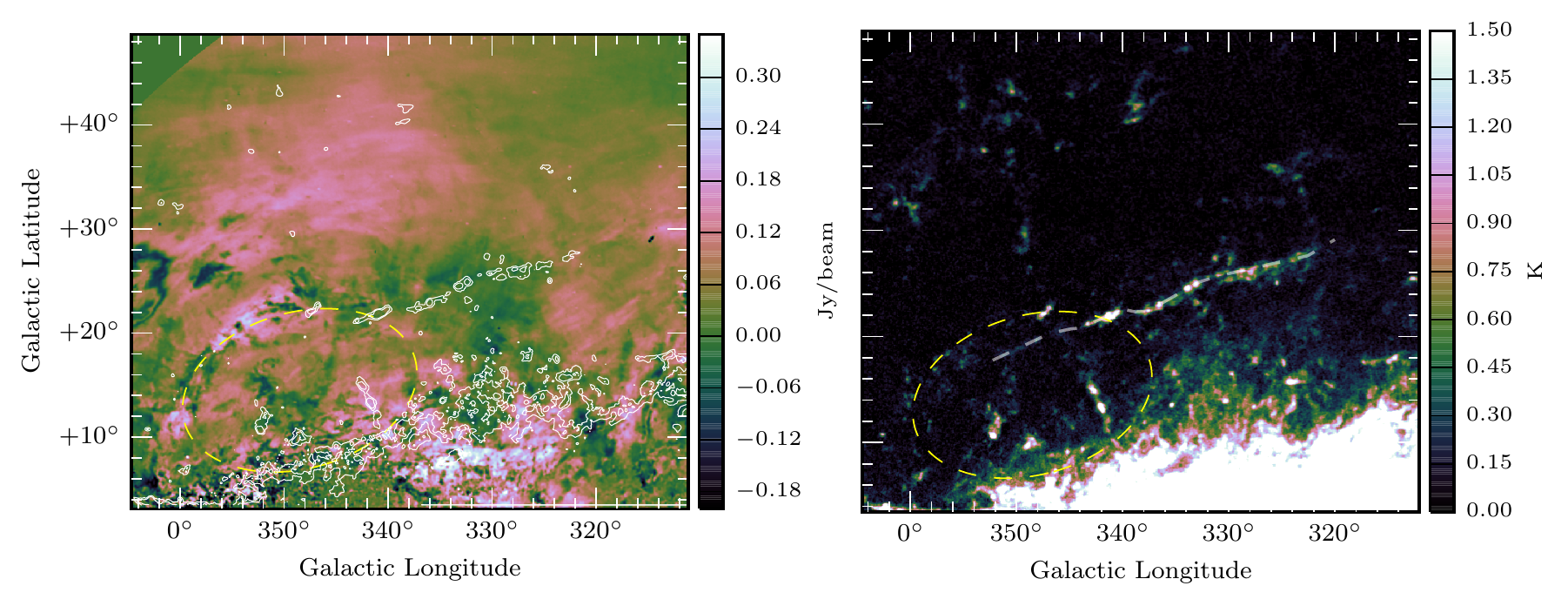}
\caption{Left panel: GASS data showing the HI emission at $V_{LSR}=-57.71$ km s$^{-1}$ with white contours. The contour levels are at $0.5$ and $1.0$ $K$. The background colour scale is the Stoke $Q$ map from S-PASS. The yellow dashed ellipse delimits the edges of the bubble as seen in Fig. \ref{fig:HI_dust_HIP}. Right panel: HI emission map at $V_{LSR}=-57.71$ km s$^{-1}$. The white dashed line represents the slice used to produce the velocity-position plot in Fig. \ref{fig:pos-vel_HVC_a}.\label{fig:HVC_Planck_sub}}
\end{figure*}

\begin{figure}
\centering
\includegraphics[width=0.5\textwidth]{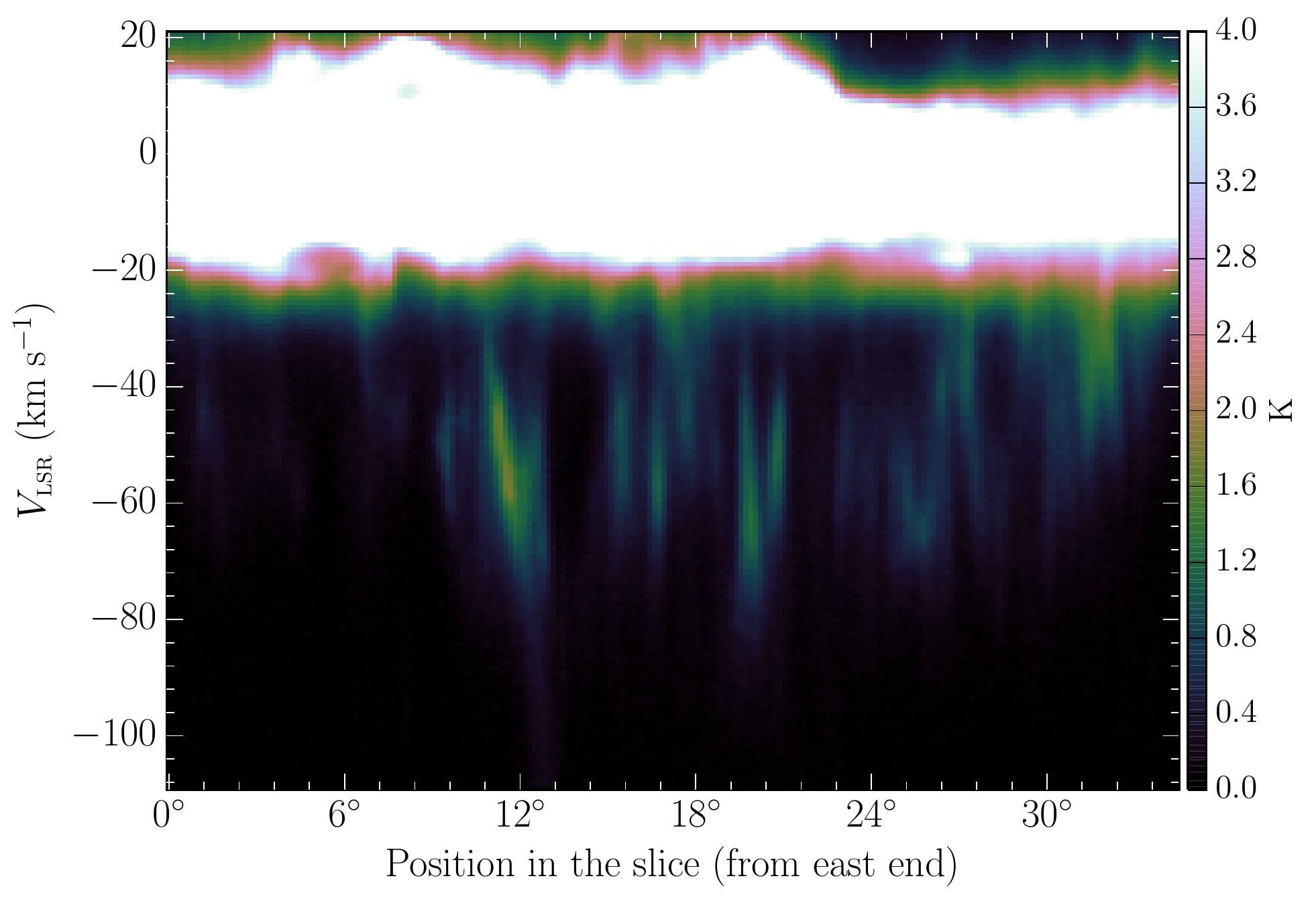}
\caption{Velocity-position plot associated with the dashed slice in the right panel of Fig. \ref{fig:HVC_Planck_sub}.\label{fig:pos-vel_HVC_a}}
\end{figure}

Another filamentary structure of HI clouds, which also seems associated with the HI cavity, is visible at $V_{LSR}=-57.71$ km s$^{-1}$. The filament is shown in Fig. \ref{fig:HVC_Planck_sub}. The brighter part of the filament starts at $l\sim343^{\circ}$, $b\sim21^{\circ}$ below the dust ridge and ends around $l\sim321^{\circ}$, $b\sim27^{\circ}$. Interestingly, this particular filament seems to have a polarised counterpart (see Fig. \ref{fig:QUP_bubble}). This spatial correspondence between the HI clumps and polarised fluctuation could be an indication that these HVCs are magnetised. The velocity-position plot associated with this filament is shown in Fig. \ref{fig:pos-vel_HVC_a}. The kinematic analysis of this filament reveals that the clumps are present along a broad range of velocities. Some of the clump velocities are extended from $V_{LSR}\lesssim-80$ km s$^{-1}$ to $\sim -30$ km s$^{-1}$. This kinematic connection between the clumps and gas at higher velocity associated with the disc could consist of material that has been stripped from the clumps by their interaction with the disc gas \citepads{2008ApJ...679L..21L, 2013ApJ...777...55H}. 

\begin{figure}
\centering
\includegraphics[width=0.5\textwidth]{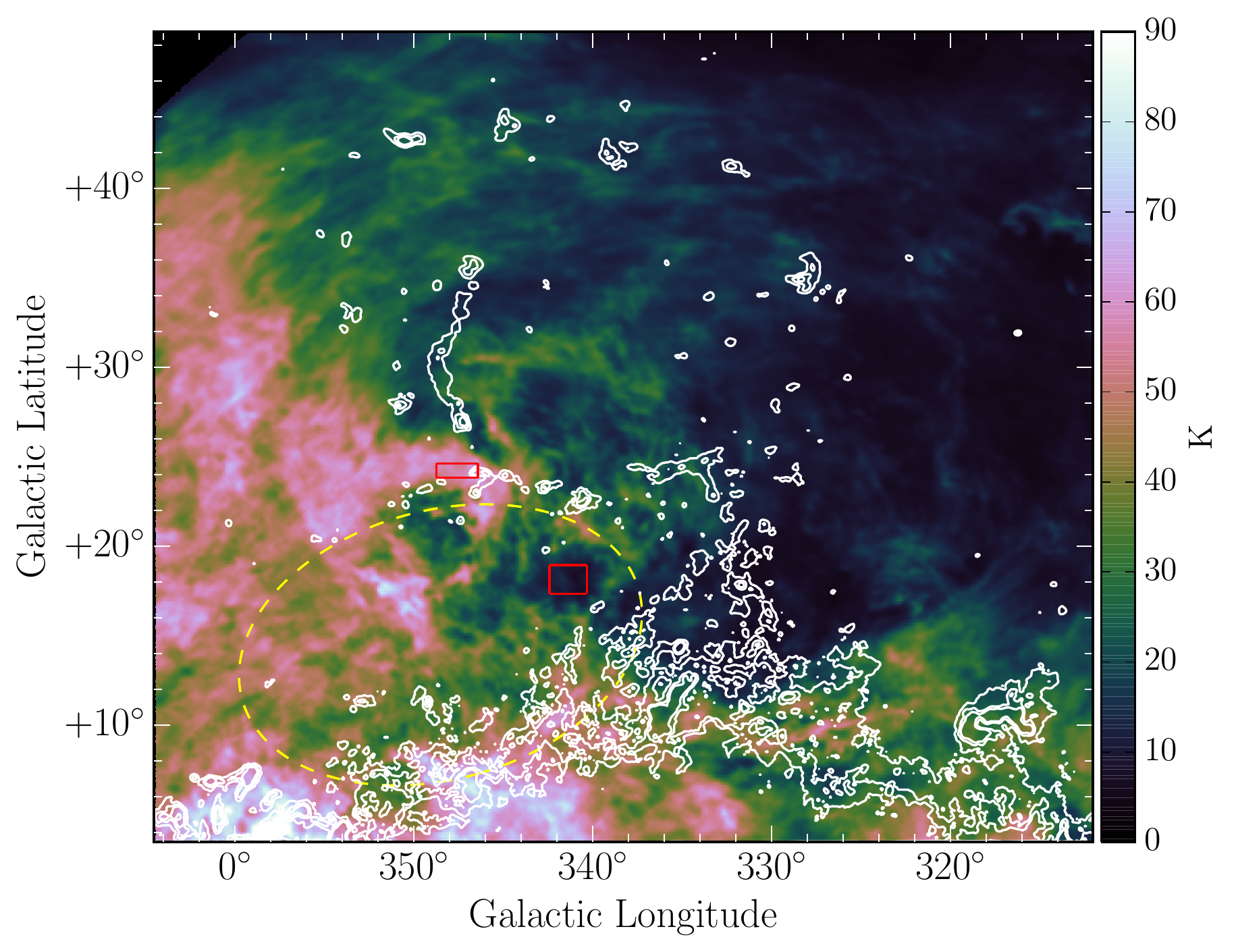}
\caption{HI line emission map at $V_{LSR}=1.65$ km s$^{-1}$ overlaid with the integrated column density map from $V_{LSR}\sim-130.27$ to $-75.03$ km s$^{-1}$. The contour levels are at 6, 14, and 22 K. The red squares mark the average regions we used to produce the spectra shown in Fig. \ref{fig:HI_spectra}.\label{fig:HVC_lowHI_overlay}}
\end{figure}

According to Fig. \ref{fig:HVC_Planck} and to the velocity-position plots in Fig. \ref{fig:pos-vel_HVC_b}, the spatial and kinematical correlation between filament II and the maximum curvature of filament I strongly suggests that the HI clumps in both filaments have the same origin. Figure \ref{fig:HVC_lowHI_overlay} shows the HI line emission map at $V_{LSR}=1.65$ km s$^{-1}$ overlaid with the integrated column density map from $V_{LSR}\sim-130.27$ to $-75.03$ km s$^{-1}$. Filament I follows the curvature of the HI ridge identified in Figures \ref{fig:HI_dust_bubble} and \ref{fig:HI_Xray_bubble} well. Similarly, filament III follows the general shape of the diffuse HI gas around $b=40^{\circ}$. It is interesting to note that the bend of filament II associated with a deceleration of the HI clumps follows the edge of the small flow identified in Fig. \ref{fig:HI_Xray_bubble} well that was also detected in X-ray. This might suggests that the deceleration of the HI clumps emerging from the HI shell that contains the hot X-ray gas is due to the interaction with the light HI flow.

The possible interaction between the HI clumps in filament II and the small flow observed at two different velocities means that the clumps have been accelerated by additional forces than just the differential pressure between the Galactic halo and inside the bubble. If the clumps are magnetised, the compressed and enhanced magnetic field on the northeastern side of the bubble might also be responsible for their acceleration. The full calculation of this mechanism is beyond the scope of this paper, but an in depth analysis of this phenomenon could lead to a radically different origin for these HVCs than the usual Galactic fountain scenario \citepads{1980ApJ...236..577B}. \citetads{2013ApJ...777...55H} showed that magnetic fields were present in the well-known Smith Cloud. If these small HI clumps were to originate inside the magnetised shell wall of the bubble, it is possible that they preserved a part of the magnetic field.

The stream of HI clumps observed at $-57.71$ km s$^{-1}$ (Fig. \ref{fig:HVC_Planck_sub}) has a different orientation compared to the three filaments at $\sim-100$ km s$^{-1}$. The velocity-position plot for this stream (Fig. \ref{fig:pos-vel_HVC_a}) also shows a different kinematic behaviour compared to the three other filaments. As mentioned before, the interaction between the stream and the disc gas could explain the kinematic connections in the velocity-position plot between the clumps and the Galactic plane. For the brightest clump, the gas velocity extends as far as $\sim100$ km s$^{-1}$. The north-east orientation of the filament is strikingly similar to the orientation of the L1709-L1755 molecular streamers associated with the Ophiuchus northernmost molecular core L1688 \citepads{1989ApJ...338..902L,1989ApJ...338..925L}. This correspondence might reveal a common origin for the Ophiuchus cloud streamers and this long stream of clumps at $V_{LSR}=-57.71$ km s$^{-1}$. This possibility is discussed further in section \ref{sec:Oph}.

\clearpage
\section{Outline of the model\label{sec:outline}}

The model presented in the previous sections, which describes a large region including the Ophiuchus and Lupus molecular complexes, agrees with the early model of Galactic chimneys by \citetads{1989ApJ...345..372N}. In this model, Galactic chimneys are locations where an upward flow of energy and magnetic flux are concentrated. Norman and Ikeuchi predicted a localised hard X-ray component associated with the hot gas streaming through the chimneys and the possible injection of HI clouds at great distances from the disc after fragmentation of the chimney walls. As suggested by \citetads{2008ApJ...688..290F}, such HI clouds would have an intermediate velocity correlated to the galactocentric radius and to the molecular cloud at the origin of the OB association. The velocities of the HI clouds presented in section \ref{sec:HVCs} do not correspond to the velocities associated with the Galactic rotation. In order to explain this discrepancy, we suggest that the morphology of the bubble combined with the compressed and enhanced magnetic field in some parts of the shell could act as a ``magnetic cannon'' and inject HI clumps at high velocity above the Galactic plane. Future works should investigate this possibility in more detail. Such magnetic cannons would challenge the origin of a part of the HVC population that is generally considered to be formed by thermal instabilities at scale heights above $\sim 1$ kpc in the hot Galactic halo gas. This hot gas stems from either an intergalactic source, according to the accretion model, or from Galactic feedback, according to the fountain model \citepads{1997ARA&A..35..217W, 2012ARA&A..50..491P}.

The following points summarise our model of the bubble structure described in the previous sections.

\begin{enumerate}

\item The bubble is located at the interface between the Galactic plane and the Galactic halo and has a regular elliptical shape that is well delimited on the western side by a subset of clouds belonging to the Lupus molecular complex (see Fig. \ref{fig:dust850}). The centre of this ellipse is located at $l,b = (348.5^{\circ},14.5^{\circ})$. According to the distance of this subset of clouds estimated by \citetads{2013A&A...558A..77G}, the centre of the bubble is at a distance of $139\pm10$ pc from the Sun. The size of the minor and major axes of the ellipse, marking the limits of the cavity, is estimated to be $36\pm4$ pc and $54\pm4$ pc, respectively.

\item This bubble is believed to expand inside a larger HI loop that is well described by \citetads{1992A&A...262..258D} (see Fig. \ref{fig:HI_dust_HIP}). This HI loop would have been created by the Upper Scorpius OB association outflows and the smaller bubble by a supernova explosion, whose progenitor was the most massive star of the association. The centroid velocity of the bubble is $V_{LSR}=1.65$ km s$^{-1}$, where the northern edge is well delimited by a HI ridge (see Fig. \ref{fig:HI_dust_bubble}).

\item The north-eastern edge of the bubble is delimited by the Ophiuchus north molecular clouds, called the Oph ridge in this model (see Figs. \ref{fig:dust850} and \ref{fig:HI_dust_bubble} or the schematic view in Fig. \ref{fig:Model}). Like the Lupus molecular complex, the Oph ridge location intersects with the boundaries of the larger HI loop at $V_{LSR}=5.77$ km s$^{-1}$ (see Fig. \ref{fig:HI_dust_HIP}). These particular locations for the two young molecular clouds, which form parts of the bubble boundaries, suggest that the two clouds were formed by multiple compressions of the local gas.

\item The bubble cavity is filled with hot gas that is well traced by its X-ray emission (see Figs. \ref{fig:X-ray_NH_bubble} and \ref{fig:HI_Xray_bubble}). The X-ray emission overshoots the western and north-western edges of the bubble, suggesting that hot gas outflows breach the cavity, possibly through the fragmented Lupus complex. Comparisons between the X-ray spatial distribution and the total gas column density map derived from Planck strongly suggest that the X-ray emission outside the cavity, notably above the northern part of the bubble, is not simply absorbed by matter along the line of sight, but that it is rather confined inside the bubble cavity.

\item HI clumps that spatially correspond well to the cavity boundaries, as well as clumps that on spatial and kinematic grounds seem to have been ejected from the cavity, have been found at LSR velocities of $\sim-96$ km s$^{-1}$ and $\sim-58$ km s$^{-1}$. The presence of these clumps at such different velocities is puzzling, and their potential origin is discussed further in section \ref{sec:HVCs_disc}.

\item Just below the Oph ridge, on the north-eastern edge of the bubble and on the eastern edge, lies the GCS, whose curvature corresponds well to this side of the cavity (see Fig. \ref{fig:QUP_bubble}). The connection between the GCS and the bubble is uncertain. The GCS has a higher polarisation intensity than the other sections of the shell, and in contrast to the other sections, here the magnetic field is aligned along the bubble boundary. This morphology of the magnetic field can be the result, similar as for the Oph ridge and Lupus, of large compressions with the expansion of the bubble. Moreover, this enhanced magnetic field could play a role in the ejection of HI clumps seen at $V_{LSR}\simeq-58$ km s$^{-1}$.

\end{enumerate}

\section{Discussion\label{sec:discussion}}

The large-scale inhomogeneous distribution of the hot gas seen through X-ray emission has previously been attributed by G15 to the asymmetric energy source of the subgroup. However, the striking correspondence between the high-intensity X-ray and the HI ridge located at $V_{LSR}=1.65$ km s$^{-1}$ (see Fig. \ref{fig:HI_Xray_bubble}) seems to indicate that a second, smaller, and probably younger, expanding HI shell is embedded in the larger expanding USco HI shell described by \citetads{1992A&A...262..258D}. Moreover, when the HI ridge is compared with the $N_{\textrm{H}}$ Planck map, it fits the morphology of an ellipsoidal bubble well. This bubble is partly completed by the Ophiuchus and Lupus molecular complexes (see Fig. \ref{fig:HI_dust_bubble}). These observations, among the others mentioned in this paper, require a new model that describes the origin of these two well-known molecular clouds and their ongoing star formation activity.

\subsection{Origin of the bubble\label{sec:origin}}

Previous studies suggested that this region has been affected by at least one supernova explosion \citepads{1989ApJ...338..902L, 1989ApJ...338..925L, 1992A&A...262..258D, 2000ApJ...544L.133H, 2001A&A...365...49H, 2001PASJ...53.1081T, 2008AstL...34..686B, 2009ApJ...700.1609M, 2010A&A...522A..51D, 2015A&A...584A..36G}. In his model, \citetads{1992A&A...262..258D} suggested that massive stars in the USco subgroup ionised the region and first contributed to the shell expansion. The energy output of the subgroup association is too low to account for the kinematics of the shell. The most massive star ($\sim$O7), which had the largest contribution to the shell expansion, probably died as a supernova, adding more kinetic energy to the expansion. \citetads{1992A&A...262..258D} suggested that this massive star was part of a binary system and that the secondary star could be the fast-moving runaway star $\zeta$ Oph that was kicked by the supernova explosion. This type of event is often called the binary-supernova scenario (BSS). According to the kinematics of $\zeta$ Oph, the runaway star could either have left the USco subgroup $\sim1$ Myr ago or have left the UCL subgroup $\sim3$ Myr ago \citepads{2000ApJ...544L.133H, 2001A&A...365...49H}. 

This hypothesis has been studied in detail by \citetads{2001A&A...365...49H}, who proposed the runaway pulsar PSR B1929+10 (J1932+1059) as the primary star of the binary system. In order to demonstrate that the two objects are the remains of the most massive binary system in the USco association, the authors calculated the past orbits of the objects and simultaneously determined their separation. The results show that a small fraction of the simulations are consistent with the hypothesis that the runaway star and the pulsar were once part of a binary system $\sim1$ Myr ago in the USco association. Later, \citetads{2004ApJ...604..339C}, with new measurements for the proper motion and parallax of B1929+10 using the NRAO Very Long Baseline Array (VLBA), reported that the latter scenario is extremely unlikely. This different conclusion is also supported by \citetads{2015A&A...577A.111K}, who performed new simulations with updated proper motion and parallax values for B1929+10 using the European VLBI Network (EVN). It is worth mentioning that two other analyses by \citetads{2008AstL...34..686B} and \citetads{2010MNRAS.402.2369T} reported that the encounter between the two objects about 1 Myr ago was still likely by increasing the errors of the parallax and the proper motion for the pulsar by a factor of 10 ( and by a factor of 30 by \citetads{2008AstL...34..686B} for the proper motion). However, since the new EVN astrometry of B1929+10 agrees well with the previous VLBA measurements by \citetads{2004ApJ...604..339C}, robust constraints on the uncertainties were retained by \citetads{2015A&A...577A.111K} for their calculations.

\begin{figure}
\centering
\includegraphics[width=0.48\textwidth]{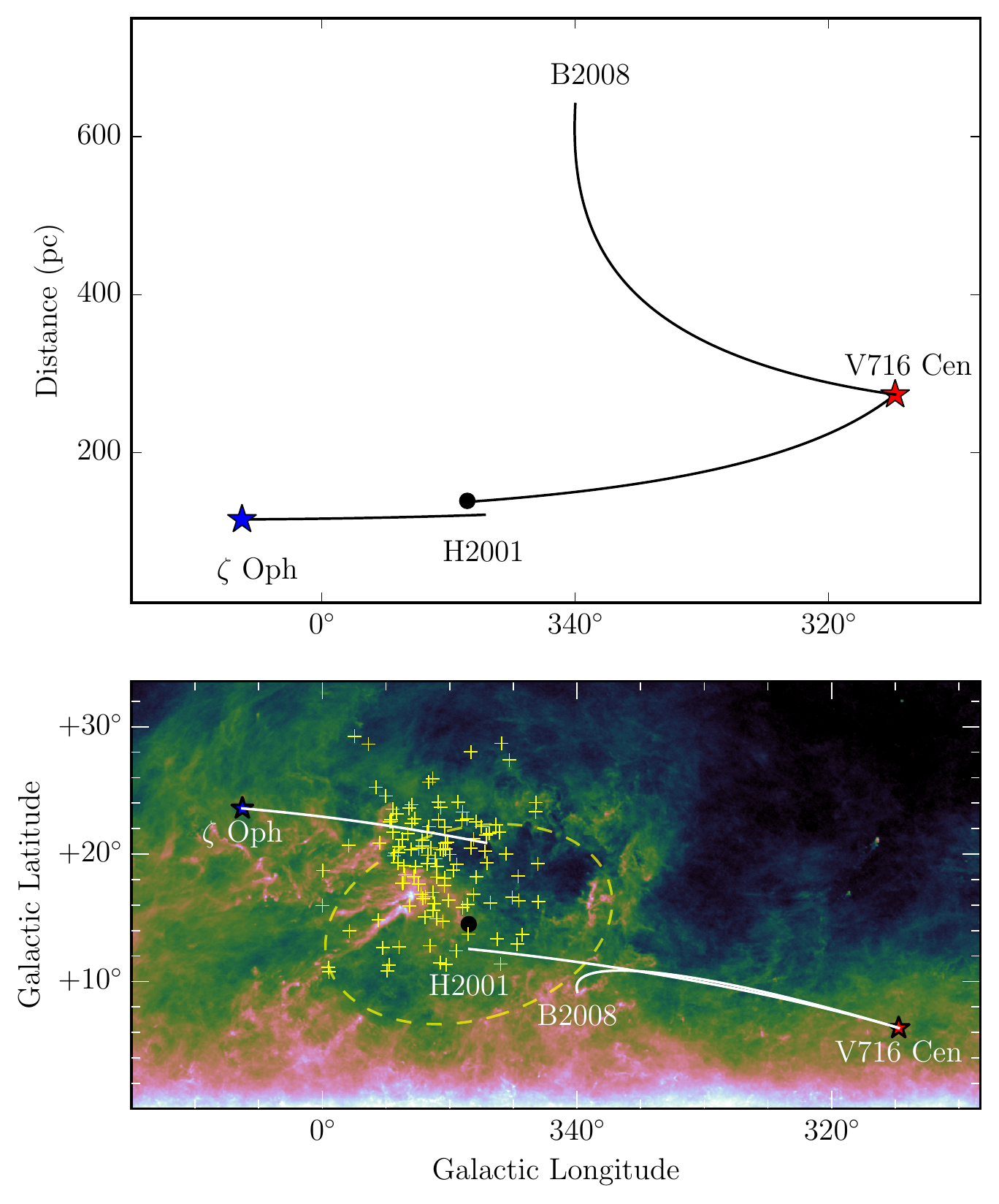}
\caption{Previous orbits of runaway stars $\zeta$ Oph and V716 Cen. Stars show the current position of the stars, yellow crosses show the position of the USco subgroup, and the black filled circle shows the geometrical centroid of the yellow dashed ellipse. The bubble centre distance in the top panel is 139 pc. Two trajectories were calculated for the binary V716 Cen, one using the radial velocity $\sim+57$ km s$^{-1}$ (H2001, 2.8 Myr) and the other using the radial velocity $-10.3$ km s$^{-1}$ (B2008, 15 Myr).\label{fig:trajectories}}
\end{figure}

In our analysis, we tested if the runaway star $\zeta$ Oph, without any association with the pulsar B1929+10, might have been located near the centre of the bubble. In this case, $\zeta$ Oph could have been part of a BSS process with an unknown companion. To calculate the orbit of the star, we used the python software galpy\footnote{http://github.com/jobovy/galpy} with the \texttt{MWPotential2014} model for the Galactic potential \citepads{2015ApJS..216...29B}. Normal distributions of the star properties were used to calculate the different possible previous orbits and the bubble distance using the assumed distance of $139\pm10$ pc. Uncertainties on measurement were set as the standard deviations for the normal distributions. The list of the runaway properties is summarised in Table \ref{tab:runaways}. The closest distance estimated for $\zeta$ Oph is $\sim14$ pc from the bubble centre located at $\sim 122$ pc around 2 Myr ago. These results make the BSS process involving this star at the origin of this bubble highly implausible (see Fig. \ref{fig:trajectories}).

One of the runaway star trajectories studied by \citetads{2001A&A...365...49H}, in addition to $\zeta$ Oph, might have been in the vicinity of the USco association in the past. V716 Centaurus (HIP 69491) is an eclipsing binary (B5V) with an orbital period of $\sim1.49$ day. Because of its binary nature, it would be highly surprising if this runaway star were the result of a supernova explosion. \citetads{2001A&A...365...49H} argued that one possibility is that the runaway binary could have been part of a stable triple system where the most massive star exploded as a supernova. Consequently, we calculated the possible past orbits of V716 Cen following the same procedure as for the runaway $\zeta$ Oph. The systemic radial velocity of the binary system is controversial, however. \citetads{2001A&A...365...49H} used a radial velocity of $+66\pm10$ km s$^{-1}$, which is the same as provided by the SIMBAD search database. \citetads{2010ApJ...721..469J} in their dynamical study of runaway stars produced through the BSS process in the Sco--Cen, used the radial velocity $-10.3\pm6.9$ km s$^{-1}$ derived by \citetads{2008MNRAS.385..381B} by averaging systemic velocities obtained from measured separated He I and Mg II lines. Figure \ref{fig:trajectories} shows the closest trajectory for a radial velocity of $\sim 57$ km s$^{-1}$ (H2001) and a bubble distance of $\sim 138$ pc. With these parameters, the star system past orbit was at $\sim 4.7$ pc around 2.8 Myr ago. The orbit using the radial velocity derived by \citetads{2008MNRAS.385..381B} (B2008) does not converge inside the bubble. Only the radial velocity $-10.3$ km s$^{-1}$ has been used to calculate the trajectory in B2008.

According to our calculations, the binary system V716 Cen is a possible candidate that could have been associated with a BSS process at the origin of the bubble discussed in this paper. Nonetheless, more measurements are needed to confirm the BSS origin of V716 Cen and its association with the Galactic bubble.

\begin{table*}
\centering
\caption{Astrometric parameters of the runaways}
\label{tab:runaways}
\begin{small}
\begin{tabular}{lcccccccc}
\hline
\hline
        Name            &       $\alpha$ (h m s)        &       $\delta$ ($^{\circ}$ $'$ $'$)    &       $l$ ($^{\circ}$)                &       $b$ ($^{\circ}$)    &       $\pi$           &       $\mu_{\alpha}$          &       $\mu_{\delta}$  &       $v_{rad}$       \\
                                &       [J2000]         &       [J2000] &       [J2000]         &       [J2000]  &       [mas]   &       [mas yr$^{-1}$] &       [mas yr$^{-1}$] &       [km s$^{-1}$]       \\
\hline
$\zeta$ Oph             &       16 37 09.54     &       $-10$ 34 01.53          &       6.28    &       $+23.59$        &       $8.91\pm0.20$   &       $15.26\pm0.26$  &       $24.79\pm0.22$  &       $-9.0\pm5.5$\tablefootmark{a}   \\
V716 Cen                        &       14 13 39.82     &       $-54$ 37 32.26           &       314.73  &       $+6.35$ &       $3.84\pm0.44$   &       $-19.220\pm0.019$&      $-11.656\pm0.021$       &       $+66\pm10$\tablefootmark{b}     \\
&&&&&&&&$-10.3\pm6.9$\tablefootmark{c}\\
\hline
\end{tabular}
\end{small}
\tablefoot{$\zeta$ Oph astrometric parameters, position, parallaxes, and proper motions ($\alpha$, $\delta$, $\pi$, $\mu_{\alpha}$, $\mu_{\delta}$) from \citetads{2007A&A...474..653V}. V716 Cen astrometric parameters are taken from \citetads{2016A&A...595A...2G}.\\
Radial velocities ($v_{rad}$) are taken from\tablefoottext{a,b}{\citetads{2007AN....328..889K}}and  \tablefoottext{c}{\citetads{2008MNRAS.385..381B}}.}
\end{table*}

\subsection{HVCs\label{sec:HVCs_disc}}

One possible explanation for the spatial correlation between the HVCs and the HI shell is that the HI clumps at high velocity were initially created by Rayleigh-Taylor instability in the shell walls seen at $V_{LSR}=1.65$ km/s. The instabilities would have `dripped' into the lighter material inside the bubble and would be then transported with the hot gas flow at the top of the bubble due to the differential pressure between the Galactic halo and inside the bubble. According to \citetads{2003ApJ...594..833M}, the growth time of these instabilities is of the order of 1 Myr, which is lower than the hypothetical  age of the supernova explosion that would have occurred in the USco association ($\sim 2.8$ Myr, see section \ref{sec:origin}). Moreover, magnetic fields would place a lower limit on the size of the instabilities. As part of the HI ridge structure, the hook located on the right-hand side of the ridge may also be due to a Rayleigh-Taylor instability developed on the collapsing shell wall through the force of the Galactic gravitation field \citepads{1988ApJ...324..776M, 2003ApJ...594..833M}.

\begin{figure}
\centering
\includegraphics[width=0.53\textwidth]{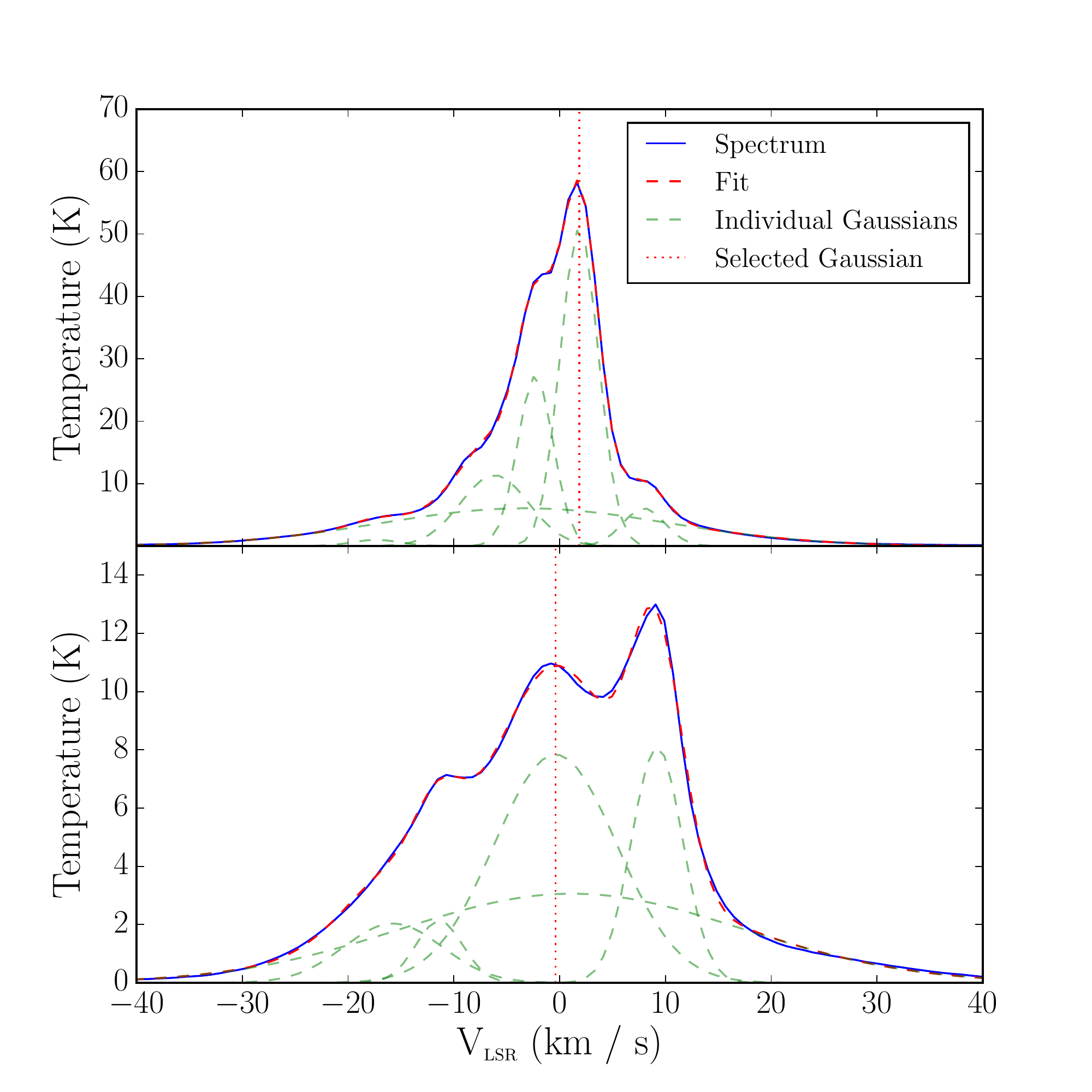}
\caption{HI spectra for the HI ridge (top) and inside the cavity (bottom). The average regions used to produce these spectra are marked by red squares in Fig. \ref{fig:HVC_lowHI_overlay}.\label{fig:HI_spectra}}
\end{figure}

The critical size, or wavelength, of the instability in the presence of a uniform magnetic field $B$ is given by \citepads{2003ApJ...594..833M}

\begin{equation}
\lambda_c=\frac{B^2\cos^2\theta}{a\mu_0(\rho_2-\rho1)}
\label{eq:R-T_critical}
,\end{equation}

\noindent where $\theta$ is the angle of the wave vector indicating the instability, $\mu_0$ is the magnetic permeability, and $\rho_1$ and $\rho_2$ are the density for the internal and shell media. In order to estimate the critical size of the instability, we estimated the densities from the HI column densities using the GASS data. The two spectra for the HI ridge and the hot internal gas are shown in Fig. \ref{fig:HI_spectra}. The average regions used to produce these spectra are marked by red squares in Fig. \ref{fig:HVC_lowHI_overlay}. The region inside the bubble was chosen for its low HI brightness at 1.65 km s$^{-1}$ and its high X-ray brightness (see Fig. \ref{fig:X-ray_NH_bubble}). However, it is also possible that the high X-ray brightness indicates that less material is in front of the bubble to obscure the X-ray emission. First, multiple Gaussians were fitted on both spectra. The Gaussian with the mean value closer to 1.65 km s$^{-1}$ was then integrated in order to calculate the column density following the relation $N_{\textrm{H}}=1.82\times10^{18}W_{\textrm{HI}}$, where $W_{\textrm{HI}}$ is the result of the integration in K km s$^{-1}$. The results of the estimated column densities are summarised and compared with the $N_{\textrm{H}}^{353}$ Planck map in Table \ref{tab:col_dens}. The derived hydrogen column densities are $4.2\times10^{20}$ and $2.1\times10^{20}$ cm$^{-2}$ for the HI ridge and the void inside the bubble, respectively. Assuming that the bubble is a symmetric ellipsoid and that its distance at the centre corresponds to the distance of the `off-cloud' stars of the Lupus complex estimated by \citetads{2013A&A...558A..77G}, then $\rho_1$ and $\rho_2$ in equation \ref{eq:R-T_critical} are $(3.0\pm0.2)\times10^{-27}$ and $(1.8\pm0.2)\times10^{-26}$ kg cm$^{-3}$, respectively. With these values, the lower limit of the size of an instability in the presence of a magnetic field with a typical value of $B\sim3\mu$G, as proposed by \citetads{2003ApJ...594..833M}, is $1.5\pm0.1$ pc. The uncertainties on these value rely only on the distance uncertainty of the bubble, and they should be considered higher. The calculation of the hydrogen column density inside the shell is also very uncertain because of the degenerate multiple Gaussian fitting and also because of the confusion along the line of sight. The density ratio $\rho_2/\rho_1$, according to the HI spectra, is $\sim 6$, which is lower than 10--20 estimated by \citetads{2003ApJ...594..833M}. However, even using $\rho_1=\rho_2/$(10--20) does not change the value of $\lambda_c$ much. Despite the large uncertainties associated with the calculation of $\lambda_c$, it is interesting to note that the critical size of $\sim1.5$ pc for the Rayleigh-Taylor instabilities corresponds well to the typical size of the HI clumps located at $-95.64$ km s$^{-1}$. This would agree with the scenario where the HI clumps are `drops' of Rayleigh-Taylor instabilities into the lighter material inside the bubble.

\begin{table}
\centering
\caption{Column densities for the HI ridge and the bubble interior (void)}
\label{tab:col_dens}
\begin{tabular}{lcc}
\hline
\hline
                                                        &               HI Ridge           &               Void                    \\
                                                        &               cm$^{-2}$               &               cm$^{-2}$               \\
\hline
Planck $N_{\textrm{H}}^{353}$           &       $1.6\times10^{21}$      &       $5.1\times10^{20}$      \\
GASS $N_{\textrm{H}}$ (total)           &       $1.2\times10^{21}$      &       $5.9\times10^{20}$      \\
GASS $N_{\textrm{H}}$ (single)  &       $4.2\times10^{20}$      &       $2.1\times10^{20}$      \\
\hline
\end{tabular}
\tablefoot{The two selected regions indicated by red squares in Fig. \ref{fig:HVC_lowHI_overlay}.}
\end{table}

\subsection{$\rho$ Ophiuchus cores and the streamers}\label{sec:Oph}

\begin{figure*}
\centering
\includegraphics[width=1.0\textwidth]{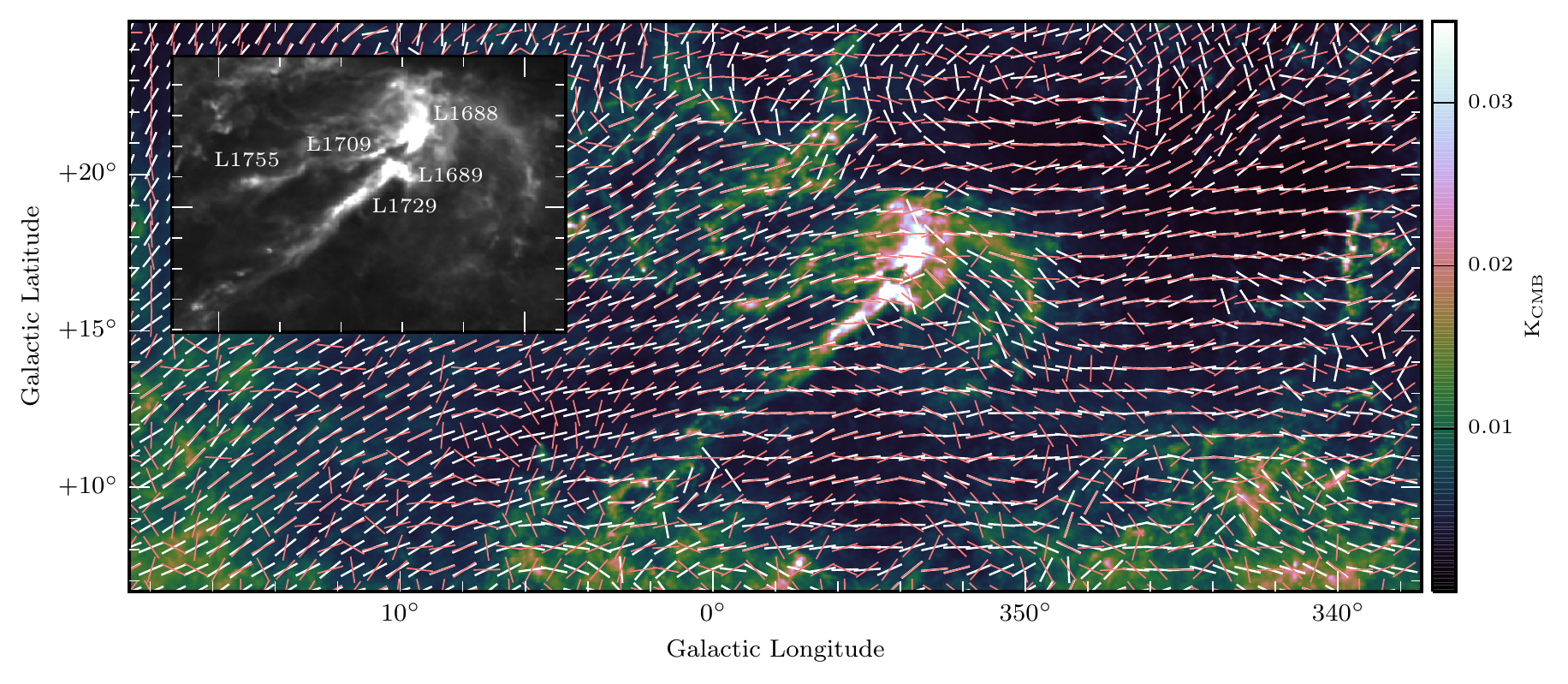}
\caption{Close-up of the the $\rho$ Ophiuchus cloud overlaid with the orientation of the magnetic field according the polarised synchrotron emission (S-PASS 2.3 GHz in red) and the orientation of the magnetic field measured with the polarised dust emission (Planck 353 GHz in white).\label{fig:Ophiuchus_pol}}
\end{figure*}

The origin of the streamers located on the southeastern side of the main $\rho$ Ophiuchus core, also called L1688, has been the subject of debates for many years. The shape of the dust arc, the $\rho$ Oph core, and the orientation of the magnetic field suggest that the core has been shocked, which could have triggered the star formation inside the core \citepads{1989ApJ...338..902L, 1992A&A...262..258D, 2001PASJ...53.1081T}. \citetads{1989ApJ...338..902L} discussed that the major star formation region in L1688 might be located in a post-shock region of the cloud. Based on the optical polarisation of background stars, the author also found evidence that the shock could have been responsible for the reorientation of the magnetic field. If a magnetised cloud is shocked, the magnetic field parallel to the shock front is amplified relative to the field perpendicular to the shock. A close-up of the $\rho$ Ophiuchus cloud overlaid with the orientation of the magnetic field according the polarised synchrotron emission (S-PASS 2.3 GHz, red) and the orientation of the magnetic field measured with the polarised dust emission (Planck 353 GHz, white) is presented in Fig. \ref{fig:Ophiuchus_pol}. The synchrotron emission at this frequency being affected by the Faraday rotation while the radiation passes through the plasma, the polarisation vectors are more suitable in this case for measuring the surrounding magnetic field orientation in the bubble. The dust polarisation is directly linked to the magnetic field of the cloud, and it is thus appropriate to measure the impact of the shock front on the magnetic field orientation of the cloud. As noted by \citetads{1989ApJ...338..902L}, the magnetic field in some part of the cloud, notably the L1688 and L1689 cores and the dust arc (indicated in Fig. \ref{fig:HI_dust_bubble}), is nearly perpendicular to the ambient field. Using the histogram of relative orientation technique applied on the $N_{\textrm{H}}$ Planck map, \citetads{2016A&A...586A.138P} also noted a drastic change in the relative orientation of the magnetic field and the filamentary structures as a function of $N_{\textrm{H}}$  for the Ophiuchus molecular cloud. The lowest $N_{\textrm{H}}$  generally has a magnetic field parallel to the structure, in contrast with a perpendicular magnetic field in the highest $N_{\textrm{H}}$ bin. For the streams, the orientation of the magnetic field measured by the dust polarisation follows the L1688-L1709-L1755 filament well, but deviates by $27^{\circ}$ in the L1729 filament. According to \citetads{1989ApJ...338..902L}, this suggests a divergence of the shock flow, where the intersection of L1709-L1755 and L1729 should locate the source of the shock. In this model, the streams were suggested to be `tongues' formed by Rayleigh-Taylor instability on the cores. Recent VLBA observations of young stellar objects in $\rho$ Oph core, L1688, measured a proper motion directed towards the dust arc \citepads{2017ApJ...834..141O}. This would also agree with the shock scenario involving the star-forming regions L1688 and L1689.

\citetads{1992A&A...262..258D} presented a detailed model where the expanding USco HI shell located at velocities $3<v<9$ km s$^{-1}$ (see Fig. \ref{fig:HI_dust_HIP}) interacted with gas of the Ophiuchus cloud that was already present in the region. He suggested the location of the near side of the expanding shell at $V_{LSR}=-12$ km s$^{-1}$, where HI structures correspond well with the contour of the molecular cloud. The interaction would result in a shock between the shell the Ophiuchus molecular cloud into which it is expanding. However, this observation is difficult to reconcile with Fig. \ref{fig:HI_dust_bubble}, where a counterpart of the dust arc is clearly associated with HI gas at $V_{LSR}=1.65$ km s$^{-1}$. This HI association with the dust arc is visible until $\sim7$ km s$^{-1}$. Thus, according to our analysis using GASS HI data, the Ophiuchus molecular cloud is more likely to interact with the far side of the expending bubble than with the near side.

According to the reorientation of the magnetic field, the two cores L1688 and L1689 are the parts of the cloud that are most affected by the shock, leaving the magnetic field associated with the streams undisturbed. \citetads{2017ApJ...834..141O} measured a distance of $137\pm1.2$ pc for the core L1688. This distance corresponds well to the distance of the bubble, $139^{+10}_{-9}$ pc, according to the off-cloud associated with Lupus. For L1689, they measured a distance of $147.3\pm3.4$ pc, suggesting that the southern streamer is about 10 pc farther away than the Ophiuchus core L1688. This new distance for the streamer challenges the scenario developed by \citetads{1992A&A...262..258D}. Assuming that the near side of the expanding shell has shocked the streamer first, it is puzzling why it would not also have changed its magnetic field orientation. The reason might be a projection effect of the stream that points towards us. The authors admitted, however, that more parallax measurements are needed to confirm the distance of L1689.

\citetads{1989ApJ...338..902L} noted that the streamers contain quasi-periodic, centrally condensed, cold dense cores instead of a uniformly decreasing material density typical of the downstream tail model. This description also fits the HI stream observed at $V_{LSR}=-57.71$ km s$^{-1}$ well (see Fig. \ref{fig:HVC_Planck_sub}). Furthermore, its orientation is similar to that of the L1709-L1755 filament. If the Ophiuchus streamers have the same origin as the HI stream seen at $V_{LSR}=-57.71$ km s$^{-1}$, they might have left the Galactic plane for the lighter hot gas in the large USco HI shell before the supernova explosion that created the smaller bubble. Then, the expansion of the second smaller shell might have interacted with the Ophiuchus cloud and compressed its gas to colder and denser gas than in the HI stream at $-57.71$ km s$^{-1}$. It is also possible that the Ophiuchus cloud left the Galactic plane after the supernova explosion and that the interaction occurred, as proposed before, between the molecular cloud and the far side of the smaller expending bubble. In either cases, the Ophiuchus molecular cloud would have a close connection with the new discovered bubble, and further analysis about this link is required.

\section{Conclusion\label{sec:conclusion}}

A careful inspection of HI data and radio polarised emission in the Ophiuchus and Lupus complex regions allowed us to identify a young Galactic bubble located at the interface between the Galactic halo and the Galactic plane. This bubble probably expands inside the larger HI shell surrounding the USco OB subgroup that has previously been identified by \citetads{1992A&A...262..258D}. Its distance is estimated at $\sim140$ pc. The bubble may have been created after a supernova explosion, for which the runaway binary system V716 Cen could be the member of an original triple system. The shock of the supernova would have swept the remaining material of the early USco molecular cloud and formed the young Lupus molecular cloud complex that is located at the western boundary of the bubble. The fragmented structure of Lupus allows some of the hot X-ray gas filling the bubble to escape at higher Galactic latitude. Parts of the Ophiuchus molecular complex, the Ophiuchus northern molecular clouds, called the Oph ridge in this model, may have the same origin.

The magnetic properties of the bubble were inspected through radio synchrotron emission at 2.3 GHz from the S-PASS data and the dust polarisation at 353 GHz from the Planck survey. The region is dominated in polarisation intensity by the GCS. This structure may be the result of shocked material following the supernova explosion at the origin of the HI cavity. The shock in this denser area of the bubble could have compressed the magnetic field and aligned its direction perpendicularly to the direction of the shock front propagation. However, other works have reported evidence that the GCS sits on the Sagittarius arm or beyond at $\sim2$ kpc or farther. We currently have not enough data to distinguish between these two possible scenarios; more data are required to conclude on this. We also suggest that the Ophiuchus molecular cloud might have a close connection to the young bubble. They are both located at similar distances, and like for the GCS, the shocked Ophiuchus cores also demonstrate that changes in magnetic field direction align perpendicularly to the shock direction.

The connection between the bubble and HVCs observed at $V_{LSR}\approx -100$ km s$^{-1}$ is one of the most surprising findings of this study. More detailed calculations and perhaps numerical simulations are necessary to confirm the ``magnetic cannon'' hypothesis, that is, if the thermal pressure and the magnetic pressure created by the expanding bubble alone might accelerate small magnetised HI clumps and launch them to high Galactic latitude. If this mechanism is confirmed, it could reveal a different scenario than the Galactic fountain model for the formation of HVCs, where HI clumps are formed at high Galactic latitude from hot gas flows from the Galactic plane.

\section*{Acknowledment}

The authors gratefully acknowledge support from the European Research Council under grant ERC-2012-StG-307215 LODESTONE. This work has been carried out in the framework of the S-band Polarization All Sky Survey (S-PASS) collaboration. The Parkes Radio Telescope is part of the Australia Telescope National Facility, which is funded by the Commonwealth of Australia for operation as a National Facility managed by CSIRO. This research made also use of Astropy, a community-developed core Python package for Astronomy \citepads{2013A&A...558A..33A} and its affiliated packages APLpy \citepads{2012ascl.soft08017R}, galpy \citepads{2015ApJS..216...29B} and pvextractor.

\bibliographystyle{apj}
\bibliography{biblio}

\end{document}